\documentclass[a4paper, amsfonts, amssymb, amsmath, reprint, twoside]{revtex4-2}
\pdfoutput=1
\usepackage[english]{babel}
\usepackage[utf8]{inputenc}
\usepackage[colorinlistoftodos, color=green!40, prependcaption]{todonotes}
\usepackage[version=3]{mhchem}
\usepackage{amsmath}
\usepackage{mathtools}
\usepackage{xcolor}
\usepackage{graphicx}
\usepackage[left=23mm,right=13mm,top=35mm,columnsep=15pt]{geometry} 
\usepackage{float}
\usepackage[most]{tcolorbox}
\usepackage{subcaption}
\usepackage{siunitx}
\usepackage{epstopdf}
\usepackage{ragged2e}
\usepackage{blindtext}

\usepackage[font=small,labelfont=bf,
   justification=Justified,
   format=plain]{caption}

\begin{document}
\title{Functionality Optimization for Singlet Fission Rate Screening in the Full-Dimensional Molecular and Intermolecular Coordinate Space }

\author{Johannes Greiner$^{a,b}$}
    \affiliation{$^{a}$Julius-Maximilians-Universität Würzburg, Center for Nanosystems Chemistry, Theodor-Boveri Weg, 97074 Würzburg, Germany.}
    \affiliation{$^{b}$Julius-Maximilians-Universität Würzburg, Institute of Physical and Theoretical Chemistry, Am Hubland, 97074 Würzburg, Germany.}
\author{Anurag Singh$^{a,b}$}
    \affiliation{$^{a}$Julius-Maximilians-Universität Würzburg, Center for Nanosystems Chemistry, Theodor-Boveri Weg, 97074 Würzburg, Germany.}
    \affiliation{$^{b}$Julius-Maximilians-Universität Würzburg, Institute of Physical and Theoretical Chemistry, Am Hubland, 97074 Würzburg, Germany.}
\author{Merle I. S. R\"ohr$^{a,b}$}
    \email[Correspondence email address: ]{ merle.roehr@uni-wuerzburg.de}
    \affiliation{$^{a}$Julius-Maximilians-Universität Würzburg, Center for Nanosystems Chemistry, Theodor-Boveri Weg, 97074 Würzburg, Germany.}
    \affiliation{$^{b}$Julius-Maximilians-Universität Würzburg, Institute of Physical and Theoretical Chemistry, Am Hubland, 97074 Würzburg, Germany.}

\date{\today} 

\begin{abstract}
In computational chemistry, accurately predicting molecular configurations that exhibit specific properties remains a critical challenge. Its intricacies become especially evident in the study of molecular aggregates, where the light-induced functionality is tied to highly structure-dependent electronic couplings between molecules. Here, we present an efficient strategy for the targeted screening of the structural space employing a "functionality optimization" technique, in which a chosen descriptor, constrained by the ground state energy expression, is optimized. The chosen algorithmic differentiation (AD) framework allows to automatically obtain gradients without its tedious implementation. We demonstrate the effectiveness of the approach by identifying Perylene Bisiimide (PBI) dimer motifs with enhanced SF rate.  Our findings reveal that certain structural modifications of the PBI monomer, such as helical twisting and bending, as well as slipped-rotated packing arrangements, can significantly increase the SF rate. 
\end{abstract}


\maketitle

\section{Introduction}
The prediction of molecular arrangements that facilitate or enhance a certain process represents a key research question in computational chemistry, spanning fields such as drug design,\cite{batool_structure-based_2019} (bio)catalysis\cite{Samha2022,nicolet_structurefunction_2020} as well as molecular electronics\cite{bamberger_beyond_2022}. Its intricacies are particularly evident in the studies of molecular aggregates, in which highly structure-dependent electronic couplings between monomers dictate the pathway of light-induced processes, leading to functionalities such as Singlet Fission (SF).\cite{bialas_perspectives_2021}The latter has garnered significant attention, as it holds the potential to exceed the theoretical efficiency limits of single-junction solar cells by yielding two triplets for the price of one photon.\cite{hanna_solar_2006,smith_singlet_2010} So far, only very few materials facilitate efficient SF, spurring extensive research aimed at identifying new potential molecular candidates\cite{akdag_search_2012,ito_molecular_2018,weber_machine-learning_2022,liu_assessing_2020,zeng_excited_2021} as well as structural arrangements \cite{wang_maximizing_2014,Renaud2013Mapping,buchanan2017singlet,singh_energetics_2021,Mirjani2014,Ryerson2019,ito_singlet_2017,farag_singlet_2018,jadhav_modulating_2022}  that enable SF. However, an unbiased screening of the full structural space requires an enormous amount of electronic structure calculations. An illustrative example of the high computational demand consists in recent work by Zaykov et al.\cite{zaykov2019singlet}: keeping the monomer structure rigid, they scanned the six-dimensional conformational space of the ethylene dimer, in order to identify the best packing motif for SF, carrying out several billion electronic structure calculations. Typically, such screening is conducted using a narrowed selection of intermolecular stacking configurations. Additionally, the rigidity assumed for the monomer structure precludes the consideration of intramolecular structural adjustments, although molecular contortion (such as twisting and bending) and the employment of nonplanar Polyaromatic Hydrocarbons has been proven as an successful approach in all fields of organic electronics.\cite{scott_geodesic_1999,ball_contorted_2015, norton_electronic_2005,sun_nonplanar_2006,guo_photoresponsive_2009,bedi_consequences_2019} In particular with respect to Perylene Bisimides, numerous experimental studies have shown how contortion leads to unique optoelectronic characteristics and further serves as a rational tool to adjust orbital energies as well as the energies of electronic transitions.\cite{liu_stringing_2020} In fact, bending of PBIs has been employed to align $S_{1}$ and $T_{1}$ energies for enhanced SF.\cite{ConradBurton2019} This aspect has also been theoretically addressed by Cruz et al., studying a series of Pyrenacenes.\cite{cruz_bending_2022}. However, the impact of contortion on SF diabatic couplings is far less explored. An alternative theoretical strategy that allows for an more efficient, targeted screening of molecular arrangements might consist in the development of  property-based optimization procedures. It is important to note that an unrestricted geometry optimization of a chosen quantity attributed to the "functionality"  is not straightforwardly applicable, as the descriptors might not be constrained by the (ground state) potential energy of the system, potentially leading to "unphysical" atomic configurations. Successful strategies consist e.g. in the development of metadynamics-based methodologies that allow for an automated extraction of structures with certain electronic properties, employing molecular dynamics simulations, in which iteratively added bias potentials steer the molecular system towards structural arrangements with high functionality, also taking into account the structural rearrangement of the individual monomers.. \cite{lindner_metafalcon_2019,lindner_metadynamics_2019, lindner_multistate_2018,rohr_new_2021}  Tom et al. have recently employed genetic algorithms along with a fitness function tailored to simultaneously optimize the SF rate and the lattice energy for the inverse design of the tetracene crystal packing, generating structures predicted to have higher SF rates while delivering insight into packing motifs associated with improved SF performance.\cite{tom_inverse_2023} 
 A promising strategy has been followed in different context by Ciminelli et al. for the optimization of minimum energy Conical Intersection (CI) structures, for which they introduced a penalty function approach, in which a modified potential energy function is minimized.\cite{ciminelli_photoisomerization_2004} These function contains two terms: the average of the state energies and a penalty function, that increases with the ground state energy. In that way, the optimization procedure targets structures with both minimized potential energy and minimized energy gap between considered states. 
In this work, we present a simple and computationally efficient functionality optimization strategy to identify molecular arrangements with enhanced SF rate  by employing a penalty function approach, for which the objective function includes an expression for the SF rate based on Fermi's golden rule augmented by an potential ground state energy dependent constraint.  
\begin{figure}[h]
    \centering
    \includegraphics[width=0.3\textwidth]{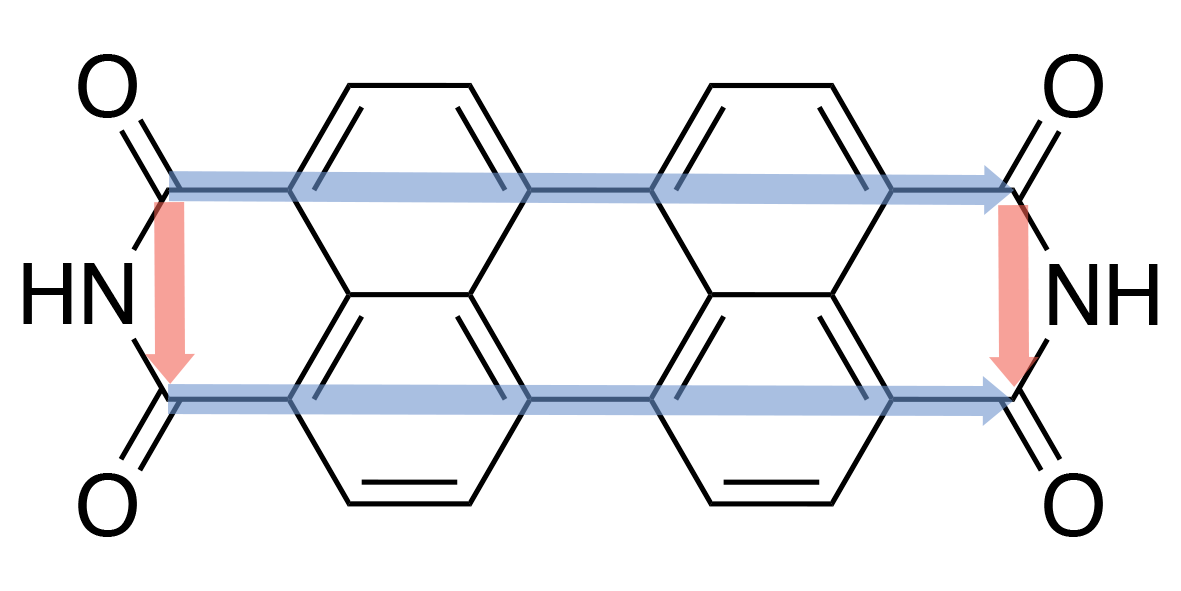}
    \caption{Structure of PBI. The carbon atoms along the blue arrows serve as guidance atoms. The mean of the blue and orange arrows define the long(\textbf{x}) and short(\textbf{y}) axis of the planar molecule. Along the atoms marked with the blue arrow the curvature is calculated.}
    \label{fig:PBI_vectors}
\end{figure}

A general challenge in the optimization of chosen functionality descriptors consist in the calculation of their gradients, as the analytic implementation can be tedious. We overcome this barrier by employing automatic differentiation (AD), that delivers gradients up to machine precision by systematically applying the chain rule to traverse the computational graph implemented functions.\cite{bartholomew-biggs_automatic_2000} While AD serves as a standard tool in machine learning, e.g. to evaluate loss functions, the implementation of quantum chemistry in the frame of the AD is an emerging field, that opens possibilities not only for gaining access to energy gradients of arbitrary order, \cite{pavosevic_automatic_2020,abbott_arbitrary-order_2021} but also in the construction and parametrization of Hamiltonians\cite{tamayo-mendoza_automatic_2018,inui_inverse_2023} as well as for inverse molecular design\cite{vargashernandez_inverse_2023}. As a consequence, the implementation of “differentiable quantum chemistry” within the AD framework is a promising direction. \cite{zhang_differentiable_2022} 
We have recently implemented model-CI Hamiltonian in the frame of AD to construct adiabatic wavefunctions in the basis of quasi-diabatic states.\cite{singh_energetics_2021} The AD directly delivered non-adiabatic couplings, which we screened as a proxy for the SF rate. 

\section{Methods}
\subsection{SF rate expression}

We describe the SF rate in the frame  of Fermi's golden rule, which is expressed as follows:

\begin{equation}
 W(\text{SF}) = \frac{2\pi}{\hbar}  T_{RP}^{2} \rho(E)
\end{equation}

Where $W(\text{SF})$ represents the rate of singlet fission, proportional to the square of the SF electronic matrix element $T_{RP}$ and a Frank-Condon weighted density of states $\rho(E)$, that depends on the energetics of the process. As our intention is not to predict total SF rates but to identify structural motifs most likely to be favorable for SF, we introduce several approximations:  
We consider only pairs of molecules, also it has been shown that polarizability  and delocalization introduced by the bulk can have significant influence on the SF process. Furthermore, we focus on the geometry optimization for maximizing the rate of the formation of the singlet-paired triplets as a key intermediate. For the expression of  $T_{RP}$ , we employ the diabatic frontier orbital approach, widely applied by the Michl group and others, that considers diabatic states formed for four electrons within the four frontier orbitals  $h_A$, $l_A$, $h_B$, $l_B$ of the two monomers $A$ and $B$ \cite{Michl2007,smith2010singlet,Buchanan2017}. 
 The orbitals are assumed to be real, normalized and orthogonal to each other. For each monomer, 3 electronic states are considered, these are: the ground state $S_0^I$, the singlet excited state $S_1^I$ and a triplet state, T$_1^I$. By removing or adding an electron to a monomer, the radical cation C$^I$ or anion A$^I$ is formed. From that, a subspace of the seven lowest electronic singlet states of the combined system can be constructed that can be classified as the ground state $^1\!(S_0^AS_0^B)$, two local excited  (LE) states $^1\!(S_1^AS_0^B)$ and $^1\!(S_0^AS_1^B)$, two charge transfer states $^1\!(\text{C}^A\text{A}^B)$ and $^1\!(\text{A}^A\text{C}^B)$ and two doubly excited states $^1\!(\text{T}_1^A\text{T}_1^B)$. The last state is commonly excluded as the energy of two localized singlet excitations is much higher than that of two localized triplet ground states paired into a singlet. 

The singlet fission rate is than assumed to be proportional to the squared matrix element of the diabatic Hamiltonian linking the exciton and the biexciton triplet states. The direct coupling between these states is negligible, while the indirect coupling through higher-level charge transfer states that couple with both the singlet exciton and the spin-correlated triplet pair state is significant.
The resulting matrix element reads as follows:
\begin{equation}
 T_{RP} =\left \vert \frac{\langle \text{LE} | \mathbf{\hat{H}} | \text{CA} \rangle \langle \text{CA} | \mathbf{\hat{H}} | \text{TT} \rangle}{\Delta E(\text{CA})}  + \frac{\langle \text{LE} | \mathbf{\hat{H}} | \text{AC} \rangle  \langle \text{AC} | \mathbf{\hat{H}} | \text{TT} \rangle }{\Delta E(\text{AC})}   \right \vert
 \label{eqn:trp}
\end{equation}

Here, $\Delta E(\text{CA})$ and $\Delta E(\text{AC})$ represent the energy differences between states CA and LE and between states AC and LE, respectively.

In the simple frontier orbital model combined with the zero differential overlap approximation, eqn. \ref{eqn:trp} reduces to

\begin{equation}
     ^{Fock}T_{RP}  = \left \vert\frac{( l_{A} | \mathbf{\hat{F}} | l_{B} ) ( l_{A} | \mathbf{\hat{F}} | h_{B} )}{\Delta E(\text{CA})} - \frac{( h_{A} | \mathbf{\hat{F}} | h_{B} ) ( h_{A} | \mathbf{F} | l_{B} ) }{\Delta E(\text{AC})}\right \vert
    \label{eqn:trp_fock_approximation}
\end{equation}
where $\mathbf{\hat{F}}$ is the Fock matrix. 
The rate expression ($T_{RP}$) was approximated by transitioning from fock matrix elements to overlap matrix elements, represented as $\langle \phi |\mathbf{\hat{F}}| \psi \rangle = \epsilon \langle \phi| \psi \rangle$. This approximation aligns with the further simplification of Michl's model \cite{buchanan2017singlet}. The value of epsilon is considered constant through the optimisation it can be omitted.

\begin{equation}
     ^{Overlap}T_{RP} = \left \vert\frac{\langle l_{A} | l_{B} \rangle \langle l_{A} | h_{B} \rangle}{\Delta E(\text{CA})} - \frac{\langle h_{A} | h_{B} \rangle \langle h_{A} | l_{B} \rangle }{\Delta E(\text{AC})}\right \vert
\end{equation}
Also $\Delta E(\text{CA})$ and $\Delta E(\text{AC})$ are approximated to be the  and are set to one. This leads to the final approximation for the singlet fission rate :
\begin{equation}
     T_{RP} = \left \vert \langle l_{A} | l_{B} \rangle \langle l_{A} | h_{B} \rangle - \langle h_{A} | h_{B} \rangle \langle h_{A} | l_{B} \rangle \right \vert
    \label{eq:sfr_overlap}
\end{equation}

\section{Computational Details}
\subsection{Function}
A simple optimization of the matrix element  $ T_{RP}$ as derived in Sec. 1.1 with respect to the nuclear coordinates will lead to unphysical structures, specifically to implosion of the molecular system, as the overlap of orbitals without any boundary is favored. However, the introduction of suitable constraints is not straight-forward. In this work, we aim to identify dimer structures with enhanced coupling, for which both the intermolecular packing motif as well as structural contortion of the monomer is allowed, while the rearrangement of atoms within molecules or formation of bonds between monomers should be unfavorable. We therefore construct a nuclear coordinate-dependent objective function, that consists of a term for the  ground state potential energy E(\textbf{x}) augmented by the expression of the matrix element  $ T_{RP}$ for which we take the logarithm to smooth the "peaked" behaviour with respect to changes in coordinates. This reads in total: 
\begin{equation}
    L(\textbf{x}) = E(\textbf{x}) - \omega\; \log( T_{RP}(\textbf{x})^2)
    \label{eq:objective_function}
\end{equation}
To balance the optimisation between energy and SF rate the free parameter $\omega$ is used. As we consider $T_{RP}$ 
to be proportional to a rate, which usually has an exponential dependence on some activation energy ( $k=A e^{\frac{E_{a}}{kT}}$ , $\log( T_{RP}(\textbf{x})^2)$ ) would be defined as an activation energy, however, its physical interpretation is not obvious. 

\subsection{Optimization Procedure}
In the following, we outline the SF rate optimization procedure applied in this work: As a first step, the monomer structure is energetically relaxed employing the AM1\cite{Dewar1985} implementation in \textit{PYSEQM} \cite{pyseqm} in conjunction with the python package \textit{\mbox{geomeTRIC}} \cite{geometric}. We then generate an ensemble of initial dimer structures by randomly  translating and rotating one monomer in relation to the other, until the lowest atom-atom distance between the monomers falls between \SI{2}{\angstrom} and \SI{3}{\angstrom}. For each dimer, we optimize  the coordinates to minimize the defined objective function in equation \ref{eq:objective_function} incorporating the implemented approximation for $ T_{RP} ^2$ (see equation \ref{eq:sfr_overlap}), and the  ground state energy given by the AM1 dimer energy adjusted for  dispersion effects\cite{grimme2006semiempirical}. The free parameter $\omega$ in equation \ref{eq:objective_function} has been chosen as  0.3 based on the evaluation of test calculations  (see ESI for details \figurename~  1 ). The AD framework of the implementation allows us to automatically obtain gradients for both the SF rate and energy as well as for $ L(\textbf{x})$. For the optimization the value of the objective function and its gradient are given to \textit{\mbox{geomeTRIC}}. The entire implementation is available on \textit{GitHub}\cite{github_johannes}.

\subsection{Dimer structure analysis}
For the analysis of the optimized dimer structure, we define a set of parameters describing the dimer geometry. These are specifically the relative translations and rotations between monomers as well as the  curvature and twisting within the monomers. For that, we define a coordinate  frame for each monomer. Therefore a set of guidance atoms is used to generate a vector along the long(\textbf{x}) and the short(\textbf{y}) axis of the original planar molecule ( blued and orange  in \figurename~\ref{fig:PBI_vectors}). Using the Gram-Schmidt orthogonalization process, the orthogonality of these vectors is ensured and the vector normal to the plane(\textbf{z}) can be calculated.

The relative translations can be calculated by projecting the vector, which connects the center of mass of both monomers on the internal coordinate system of a monomer. Relative rotations are handled in a similar way. As an example, z-rotation can be determined by projection of the x-vector of one molecule on the xy-plane of the other molecule and calculation of the angle between the projected vector and the x-vector. 
curvature is calculated by projecting the guidance atoms of each longside (blue arrows in \figurename~  \ref{fig:PBI_vectors}) onto the x-z plane of the monomer's coordinate frame and fitting these points to the parabola $z(x) = a x^2 + b x + c$. The curvature, defined as the second derivative of z, is then $2a$. Since both sides are fitted independently, the mean curvature of a monomer is given.
The twisting angle of a monomer is defined as the angle between the two short axes vectors (orange arrows in \figurename~ \ref{fig:PBI_vectors}) of the molecule.

\subsection{Quantum Chemical calculations}
Energies of the ground state, $S_{1}$ and $T_{1}$ states of monomers were calculated using CASPT2 \cite{Andersson1992} employing the BAGEL software\cite{Shiozaki2018BAGEL}. The active space comprised four orbitals: HOMO-1, HOMO, LUMO, and LUMO+1, accommodating four electrons. The cc-pVDZ basis set was utilized for these computations.

\section{Results}

\begin{figure}[ht]
    \centering
    \includegraphics[width=0.45\textwidth]{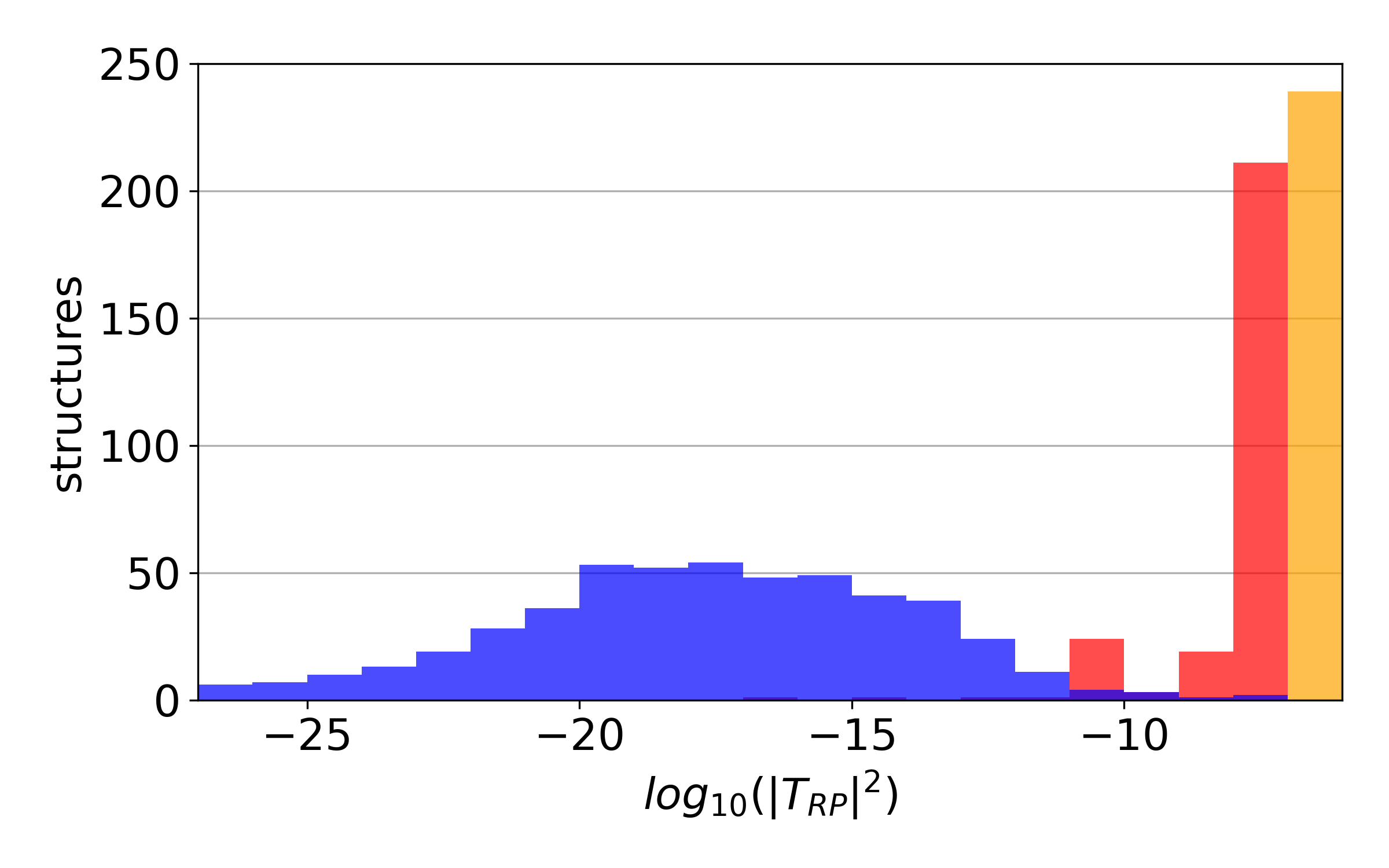}
    \caption{The red/orange histogram shows the distribution of the SF rate for the optimized structures and the blue one shows the distribution of the SF rate for the start structures. The numbers are the structure counts of the bars. The optimisation starts with a broad distribution of SF rates and can improve them to a sharp distribution of higher SF rates.}
    \label{fig:histogram}
\end{figure}

We expect the "SF rate landscape" of the PBI dimer to have several local extrema. In order to search for pronounced minima, we generate an ensemble of 500 dimer structures which we individually optimized, according to the described procedure. The resulting optimized structures are sorted after the best SF rate and analyzed in terms of the defined structural parameters. An extensive list of all  structures is given in the SI. 14 out of the 500 optimizations did not converge within the chosen limit of 300 steps. Optimiziation convergence was reached in average within \num{88} steps.
In \figurename~\ref{fig:histogram} the SF rate optimization results for the ensemble are visualized. The blue broad distribution located around \num{5.58e-11} depicts the SF rates of the initially generated dimer ensemble, while the orange sharp distribution around \num{1.20e-07} gives the optimized values, clearly illustrating the drastically enhanced SF rates of the final structures.

\begin{figure}[h]
     \centering
     \begin{subfigure}[b]{0.23\textwidth}
         \centering
         \includegraphics[width=\textwidth]{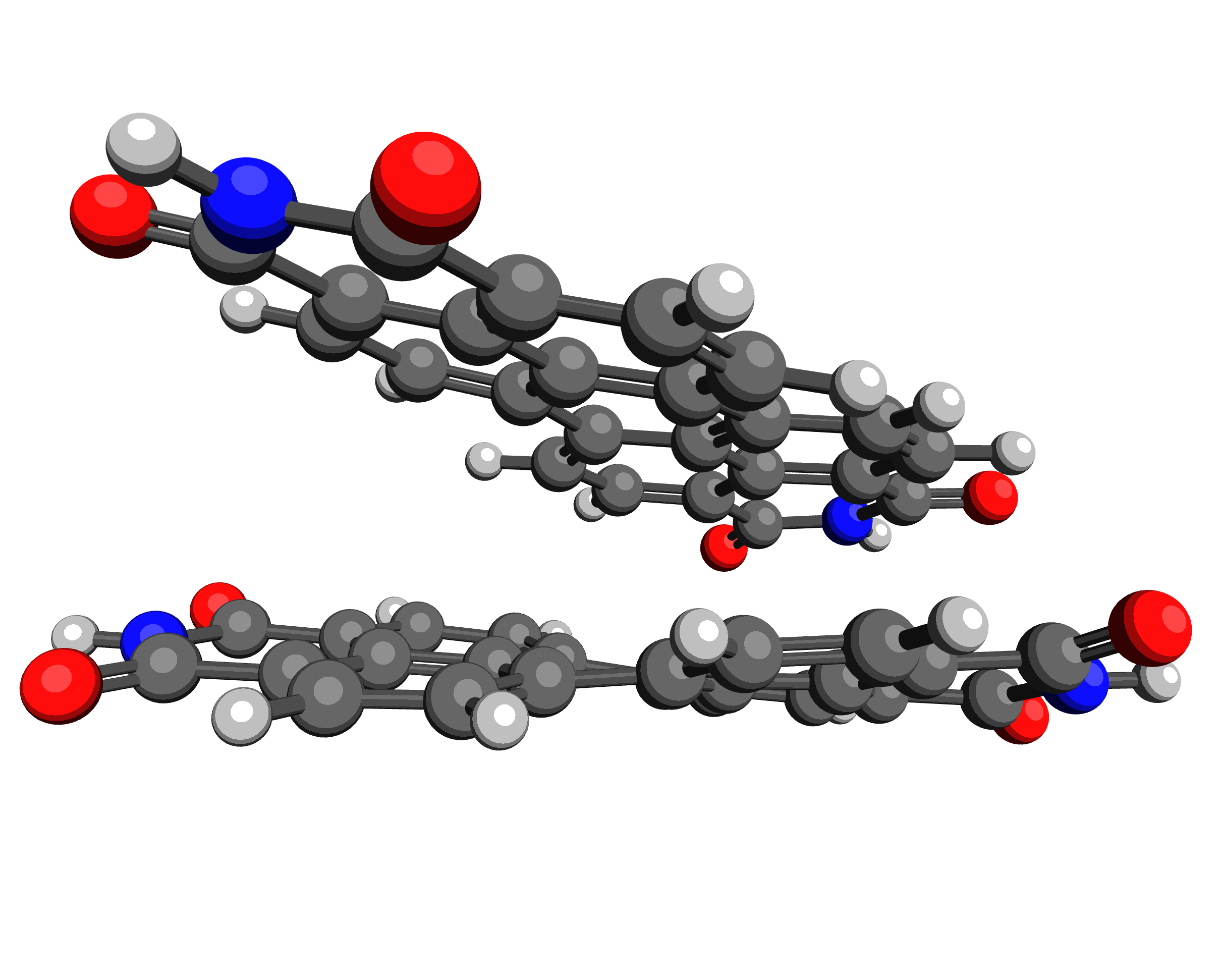}
         \caption{Side view.}
         \label{fig:struct_best_a}
     \end{subfigure}
     \hfill
     \begin{subfigure}[b]{0.23\textwidth}
         \centering
         \includegraphics[width=\textwidth]{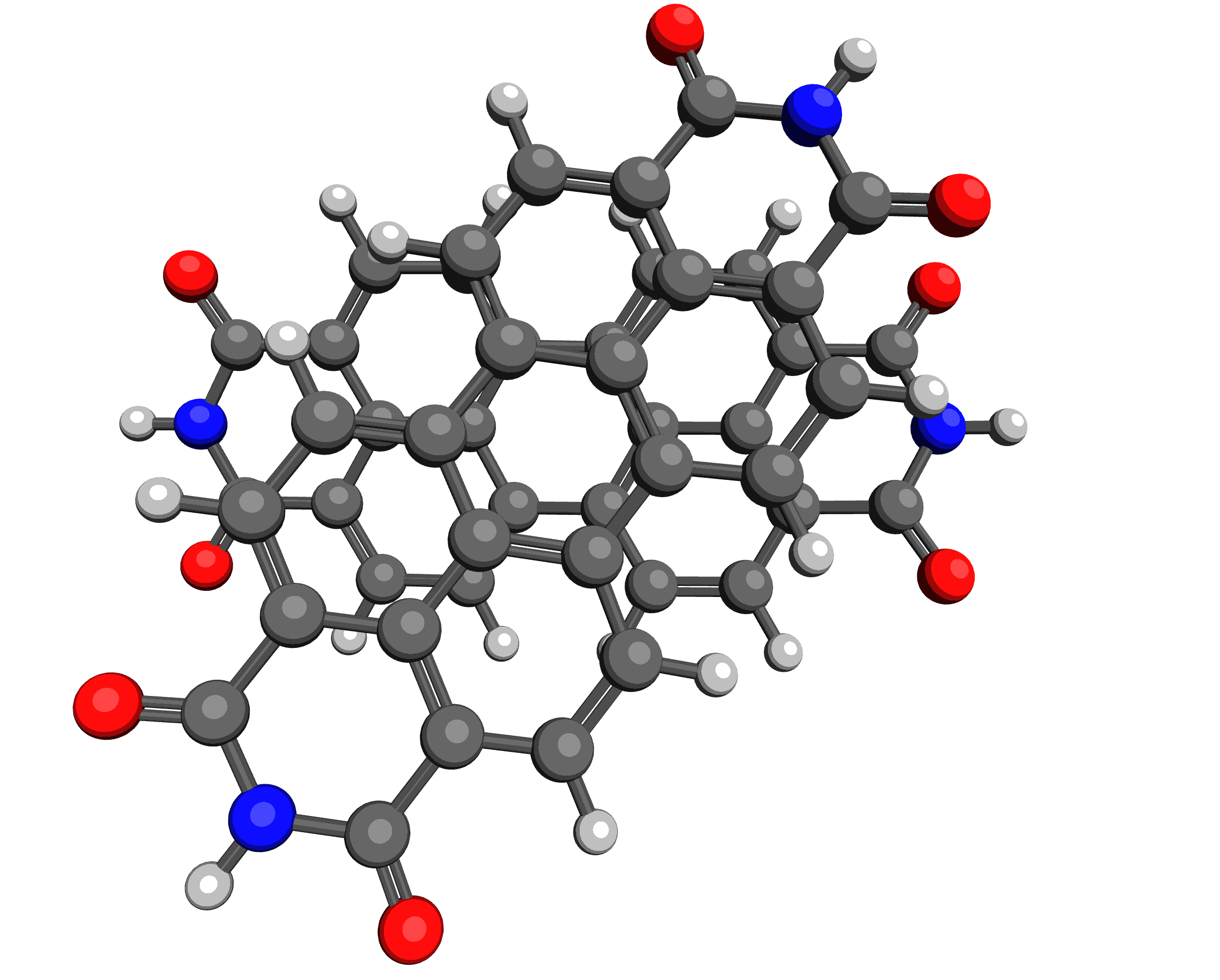}
         \caption{Top view.}
         \label{fig:struct_best_b}
     \end{subfigure}
    \caption{Top structure of the 500 trajectories with high SF rate of \num{4.06e-7}.}
    \label{fig:struct_best}
\end{figure}

The structure with a overall highest SF rate  of \num{4.06e-7}  (illustrated  in \figurename~ \ref{fig:struct_best}) converged within 51 optimization steps.   The packing motif exhibits a z-translation of \SI{3.07}{\angstrom} paired with small x- and y-translations of \SI{-0.40}{\angstrom} and \SI{-0.67}{\angstrom}, respectively.

Furthermore the monomers show a z-rotation of \SI{53.8}{\degree} with respect to each other, while the x- and y-rotations are small with \SI{6.1}{\degree} and \SI{6.9}{\degree}, respectively. Inspection of the individual monomer structures reveal, that both PBI molecules feature significant twisting angles of \SI{24.3}{\degree} (monomer $A$) and \SI{15.7}{\degree} (monomer $B$). Furthermore, monomer $A$ exhibit a noticeable curvature of \SI{10.6e-3}{\per\angstrom\squared} whereas monomer $B$ is only very slightly bent of \SI{-0.7e-3}{\per\angstrom\squared}.

To gain a deeper understanding of the structural motifs that feature high SF rates, we conducted a detailed analysis on a subset of structures with $T_{RP}$ greater than \num{1e-7}. This subset comprises 239 structures, which represent \SI{47.8}{\percent} of all optimized structures and is indicated by the orange bar in \figurename~\ref{fig:histogram}. The SF rate ranges from \num{1.0e-07} to  \num{4.1e-07} with an average rate of \num{2.0e-07}. Given that the average rotations around the x- and y-axes are relatively low ( \SI{7.1}{\degree} and \SI{6.2}{\degree}, respectively) as well as the z- axis translation ( averaging \SI{2.81}{\angstrom}) we have excluded these parameters for further analysis. Considering the symmetry of the PBI molecule, we adjusted the structural coordinates to ensure that the translations along the x- and y-axes are always positive, while the rotation about the z-axis is modified accordingly. We normalized the selected parameters for structural evaluation and applied a three-dimensional Principal Component Analysis (PCA)\cite{pca-algo,scikit-learn} to organize and reduce the dimensionality of the data. This PCA approach captured \SI{68}{\percent}  of the variance (see \tablename~\ref{tab:PCA_contributiuons}), indicating that the three PCA dimensions likely encompass the most significant structural features.

\begin{table}[ht]
    \centering
    \begin{tabular}{c|c}
        dimension & contribution \\ 
        \hline
        1 & \SI{33.05}{\percent} \\
        2 & \SI{21.72}{\percent} \\
        3 & \SI{13.26}{\percent} \\
        \hline\hline
        $\Sigma$ & \SI{68.03}{\percent}
    \end{tabular}
    \caption{Variance Contribution per dimension in PCA.}
    \label{tab:PCA_contributiuons}
\end{table}

\begin{figure}[ht]
    \centering
    \includegraphics[width=0.45\textwidth]{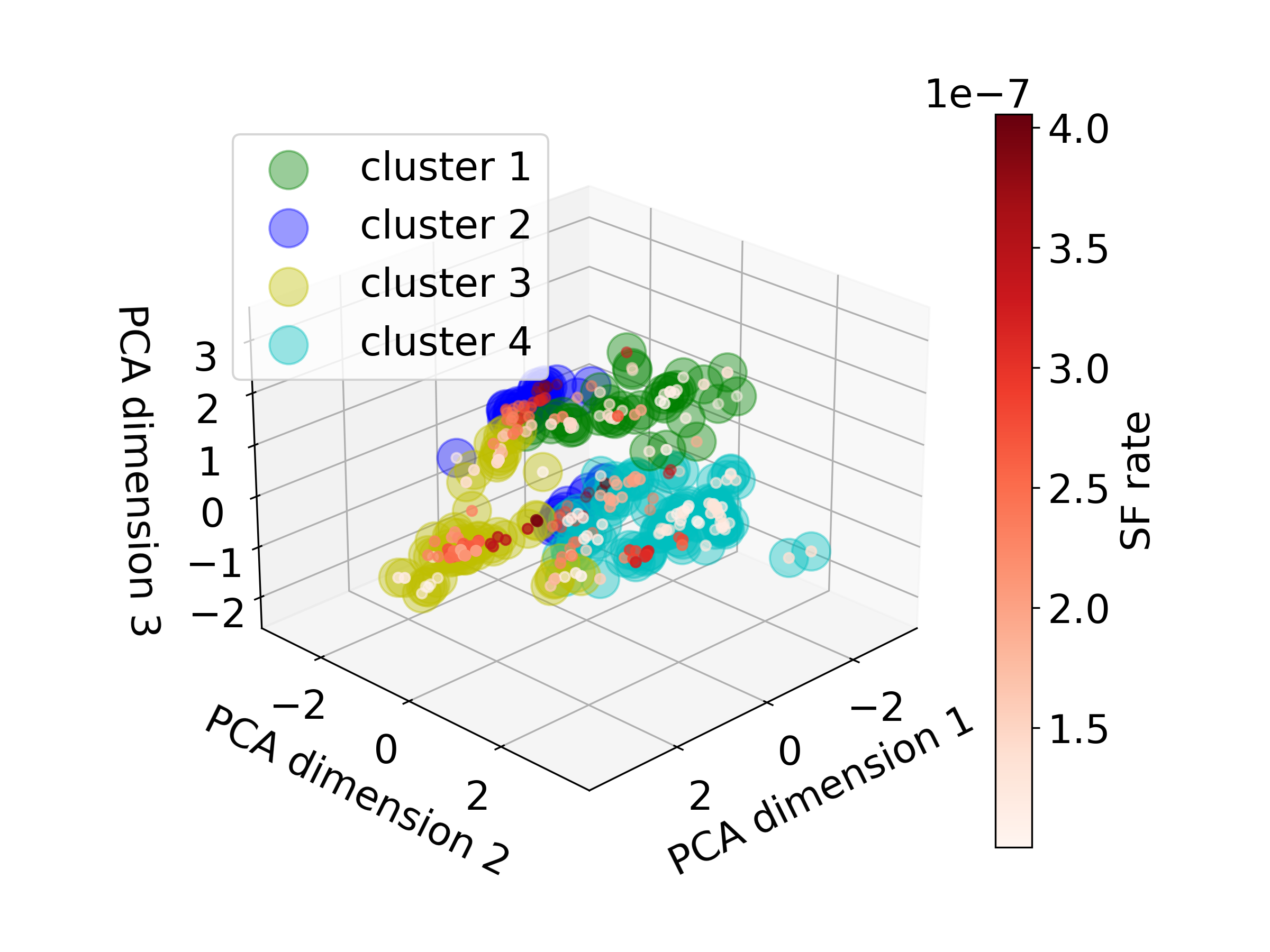}
    \caption{Three-dimensional PCA analysis of the seven attributes: x- and y-translation, z-rotation and twist angle and curvature for both monomers. The inner white to red sphere indicates the SF rate for each geometry. The SF rate scale is shown in the color bar on the right. The outer transparent sphere indicates the corresponding cluster affiliation.}
    \label{fig:3D_cluster}
\end{figure}

\begin{figure}[ht]
    \centering
    \includegraphics[width=0.45\textwidth]{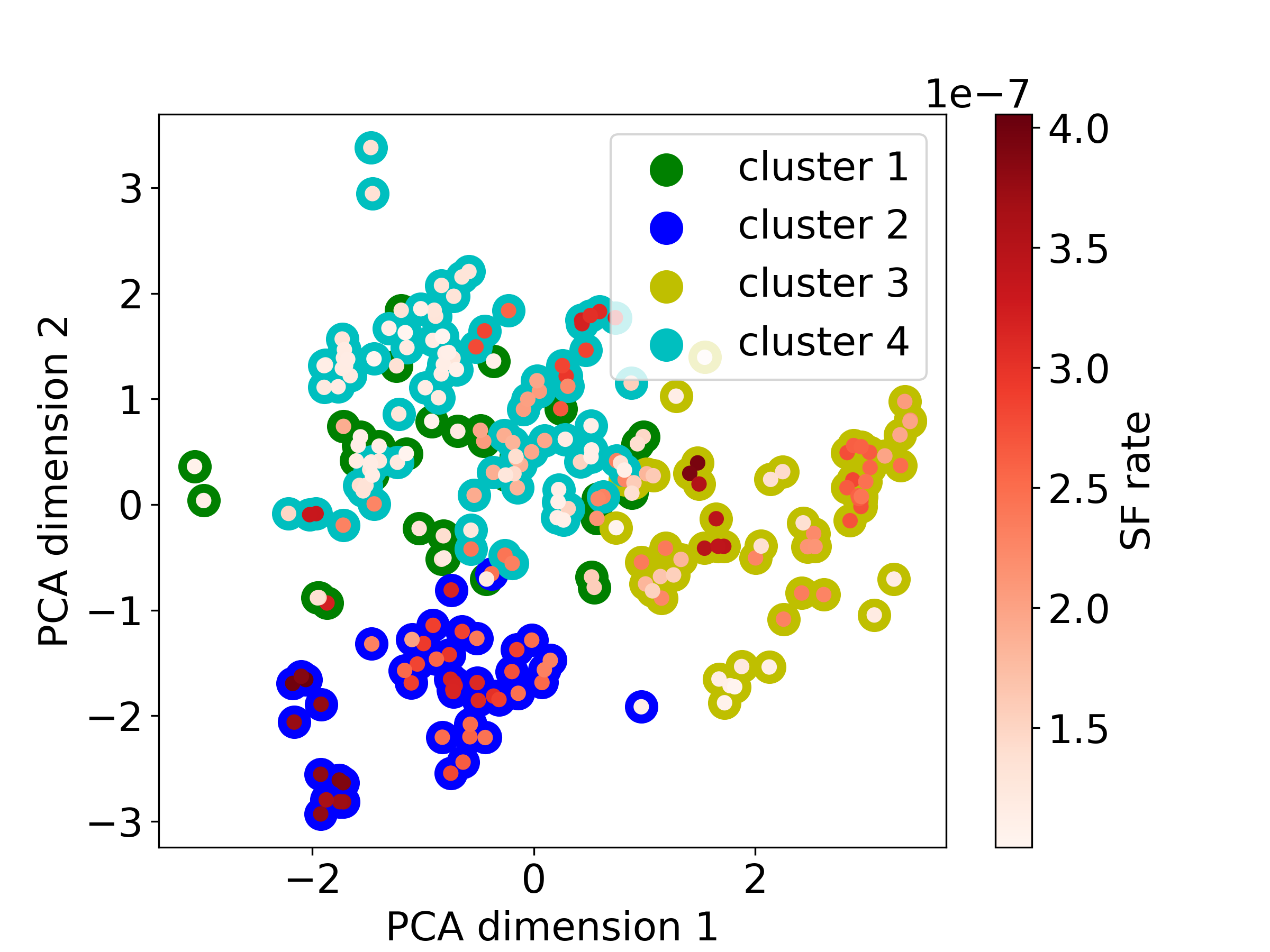}
    \caption{Two-dimensional projection onto the most significant PCA dimensions of three-dimensional \figurename~\ref{fig:3D_cluster}. The inner white to red sphere indicates the SF rate for each geometry. The SF rate scale is shown in the color bar on the right. The outer transparent sphere indicates the corresponding cluster affiliation}
    \label{fig:2D_cluster}
\end{figure}

Within the explored three-dimensional space, we identified four distinct clusters through the application of the k-means algorithm\cite{kmeans-algo,scikit-learn}, as visualized Figures in \figurename~\ref{fig:3D_cluster} and \figurename~\ref{fig:2D_cluster} with SF rates visualized using a color map.

\begin{table}[ht]
    \centering
    \begin{tabular}{|c|c|c|c|}
    \hline
        Name& $\Delta E_{T_{1}}$ in $eV$ & $\Delta E_{S_{1}}$ in $eV$ & $2 \Delta E_{T_{1}} - \Delta E_{S_{1}}$ in $eV$\\
        \hline
        \textbf{PBI-C1}$_A$    & 1.1038 & 2.2758  & -0.0682 \\
        \textbf{PBI-C1}$_B$    & 1.1068 & 2.2799  & -0.0663 \\
        \textbf{PBI-C2}$_A$    & 1.0871 & 2.2539  & -0.0795 \\
        \textbf{PBI-C2}$_B$    & 1.0989 & 2.2693  & -0.0714 \\
        \textbf{PBI-C3}$_A$    & 1.2331 & 2.1669  & 0.2993 \\
        \textbf{PBI-C3}$_B$    & 1.2332 & 2.1662  & 0.3002 \\
        \textbf{PBI-C4}$_A$    & 1.1048 & 2.2773  & -0.0676 \\
        \textbf{PBI-C4}$_B$    & 1.2274 & 2.1642  & 0.2905 \\
     
        PBI-Planar& 1.4022 & 2.4891  & 0.3154 \\
        \hline
    \end{tabular}
    \caption{$\Delta E_{S_{1}}$ and $\Delta E_{T_{1}}$ by CASPT2/cc-pVDZ(4,4)}
    \label{tab:S_1_T_1_CASPT2}
\end{table}

We further categorized the individual clusters based on their distinct structural patterns. For each cluster, we determined the mean and standard deviation of the attributes in PCA space, subsequently converting these back to the real attribute space. This approach enabled the generation of a representative "mean"-structure for each cluster, fixing the z-translation at \SI{3.00}{\angstrom}. These structures are depicted in \figurename~\ref{fig:clusters} from both top and side view.

\newcommand{\heightforbox}{4 cm}

\definecolor{plt-green}{RGB}{0,128,0}
\definecolor{plt-blue}{RGB}{25,0,255}
\definecolor{plt-yellow}{RGB}{191,191,1}
\definecolor{plt-cyan}{RGB}{3,191,191}

\begin{figure*}[ht]
\begin{center}
     \begin{subfigure}[b]{0.47\textwidth}
         \centering
         \begin{tcolorbox}[title=(a), enhanced,attach boxed title to top left={yshift=-4.0mm,yshifttext=-1mm},
  colback=white,colframe=plt-green,colbacktitle=plt-green,
  fonttitle=\bfseries,
  boxed title style={size=small,colframe=plt-green}, overlay={
            \node[inner sep=0pt,outer sep=0pt,anchor=north east] at ([xshift=-0.1cm,yshift=0cm]frame.north east) {\includegraphics[width=1cm]{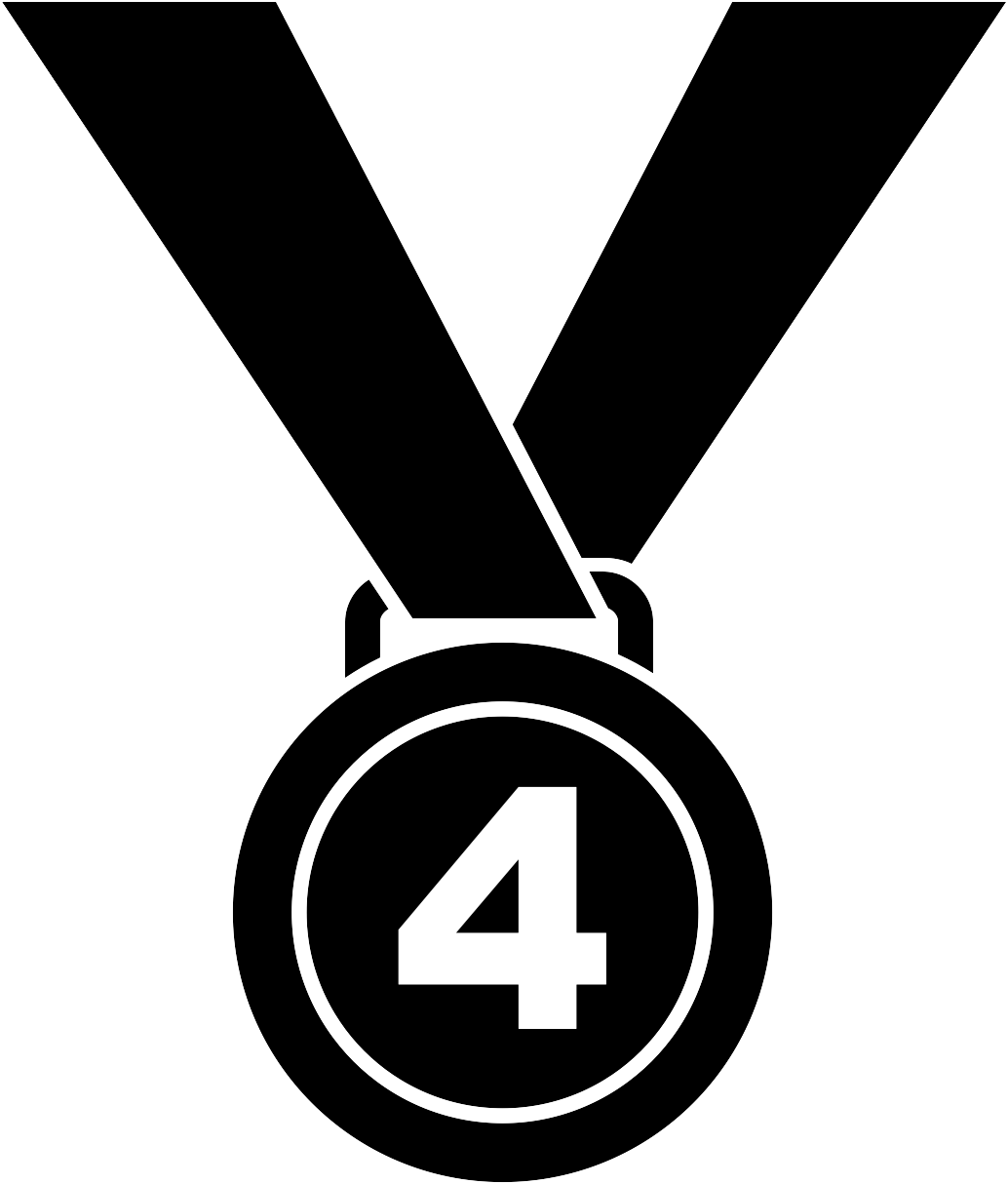}};}]
  \vspace{2mm}
            \includegraphics[width=0.49\textwidth,height=\heightforbox, keepaspectratio]{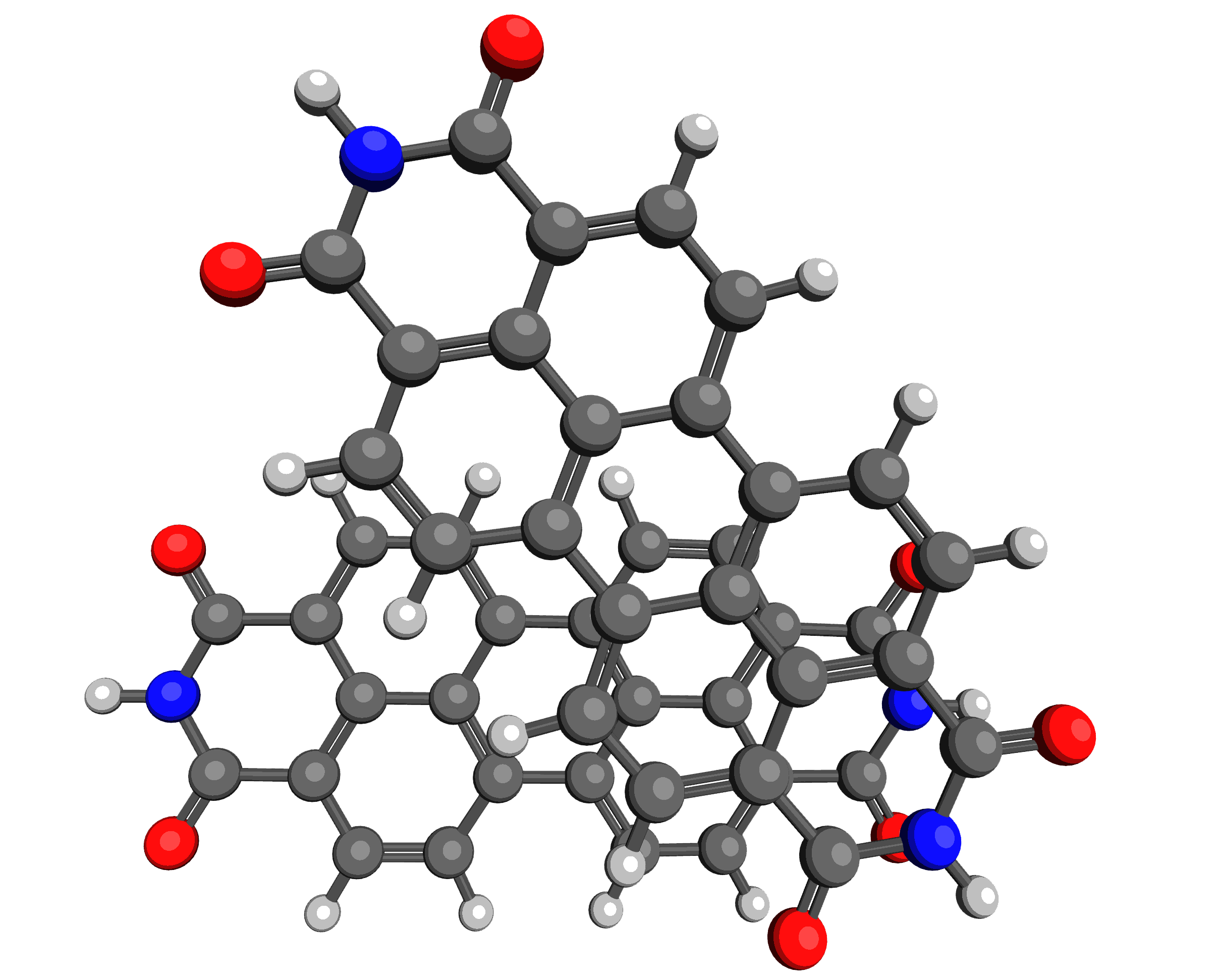}
         \includegraphics[width=0.49\textwidth,height=\heightforbox, keepaspectratio]{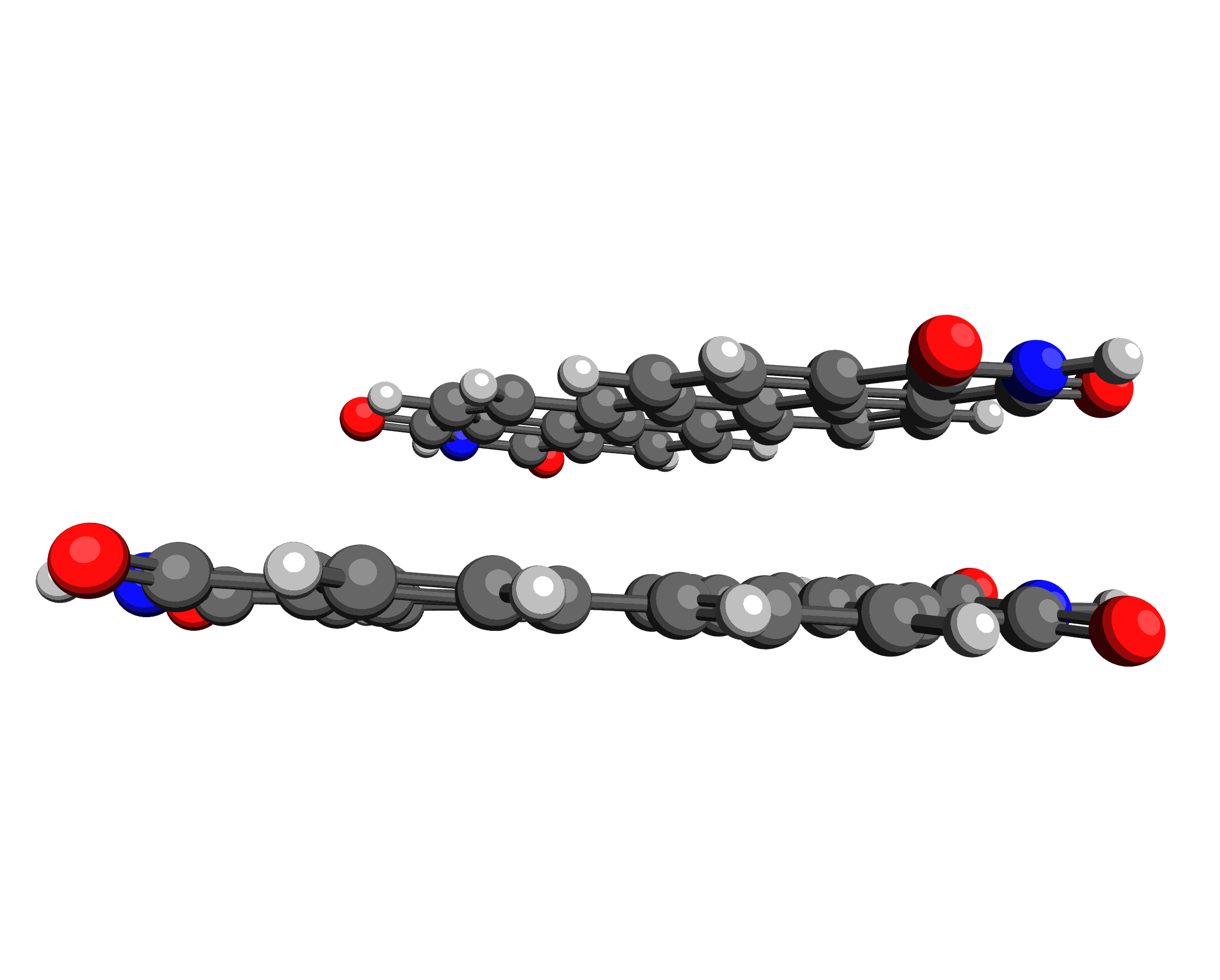}
        \end{tcolorbox}
     \end{subfigure}
     \hspace{0.3cm}
     \vspace{0.3cm}
     \begin{subfigure}[b]{0.47\textwidth}
         \centering
         \begin{tcolorbox}[title=(b), enhanced,attach boxed title to top left={yshift=-5.5mm,yshifttext=-1mm},
  colback=white,colframe=plt-blue,colbacktitle=plt-blue,
  fonttitle=\bfseries,
  boxed title style={size=small,colframe=plt-blue}, overlay={
            \node[inner sep=0pt,outer sep=0pt,anchor=north east] at ([xshift=-0.1cm,yshift=0cm]frame.north east) {\includegraphics[width=1cm]{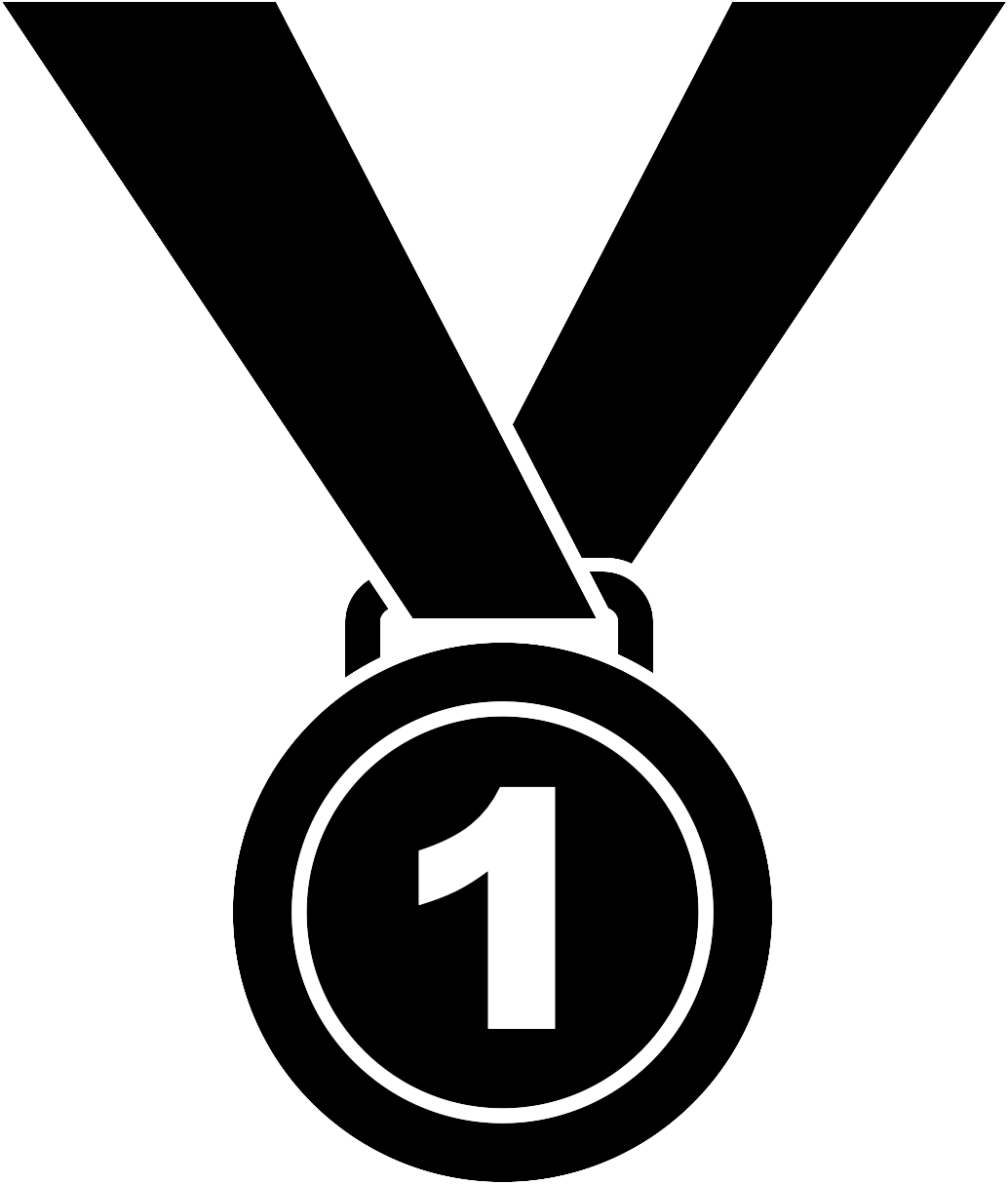}};}]
  \vspace{2mm}
         \includegraphics[width=0.45\textwidth,height=\heightforbox, keepaspectratio]{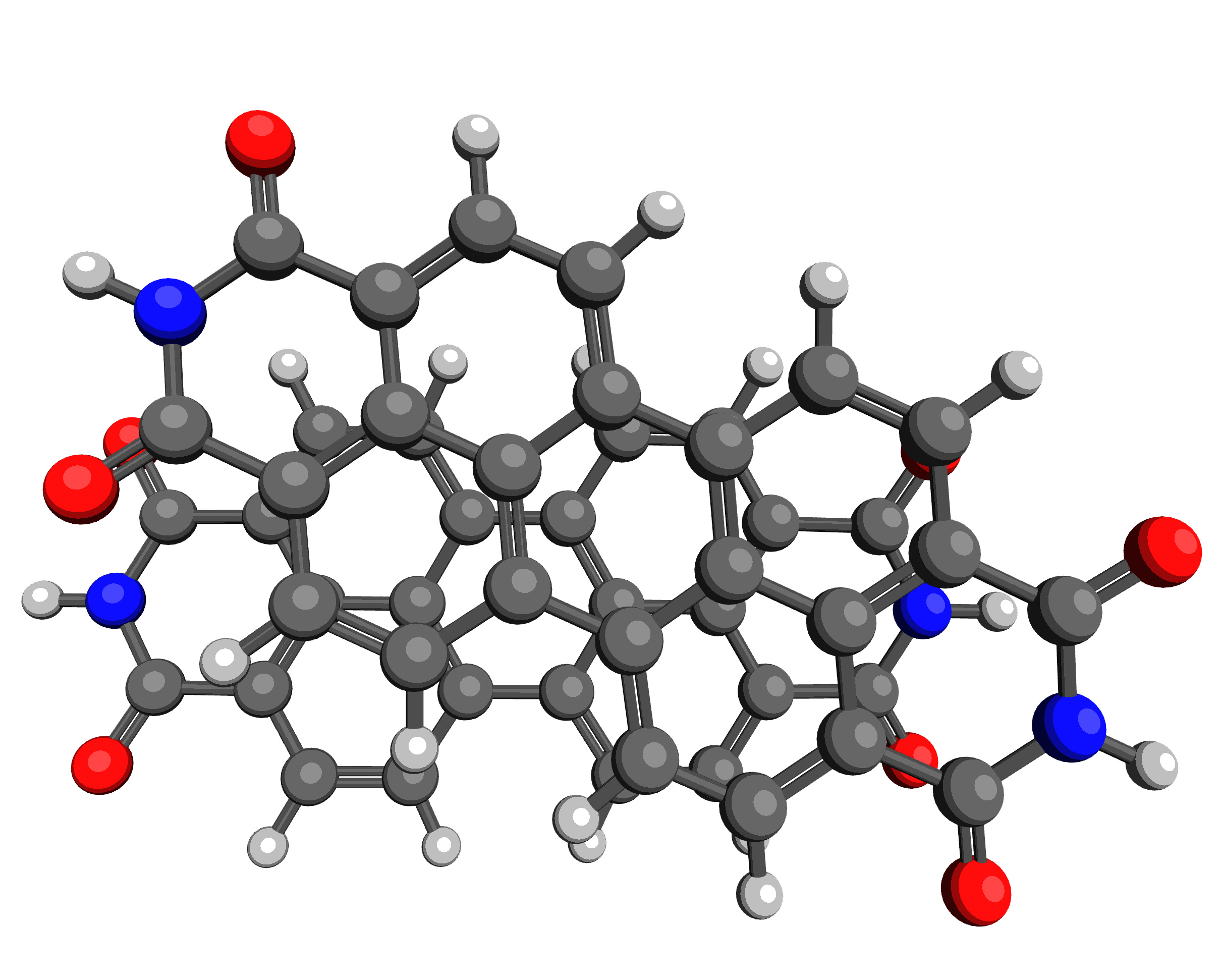}
         \hspace{2mm}
         \includegraphics[width=0.49\textwidth,height=\heightforbox, keepaspectratio]{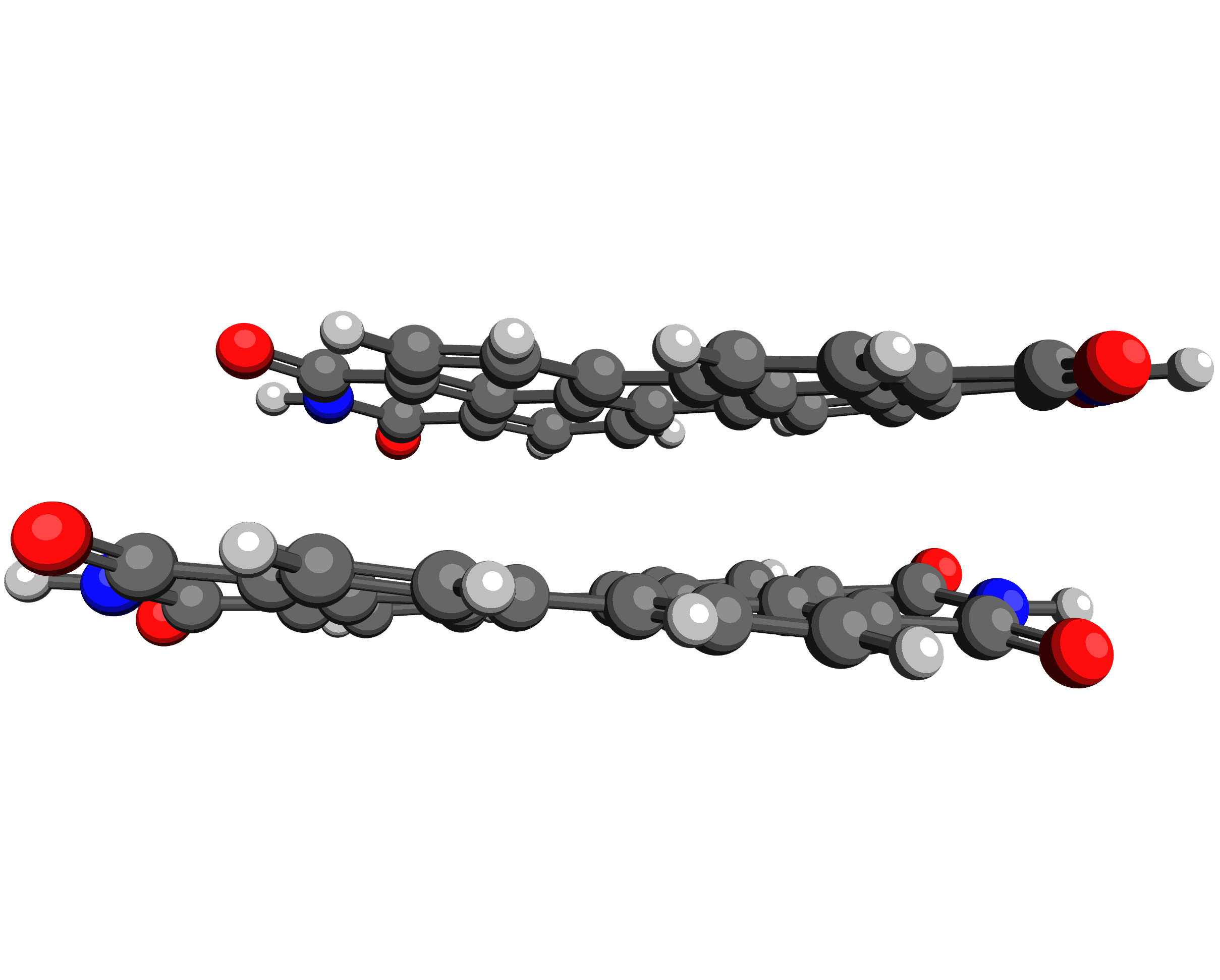}
         \end{tcolorbox}
     \end{subfigure}
     \begin{subfigure}[b]{0.47\textwidth}
         \centering
         \begin{tcolorbox}[title=(c), enhanced,attach boxed title to top left={yshift=-5.5mm,yshifttext=-1mm},
  colback=white,colframe=plt-yellow,colbacktitle=plt-yellow,
  fonttitle=\bfseries,
  boxed title style={size=small,colframe=plt-yellow}, overlay={
            \node[inner sep=0pt,outer sep=0pt,anchor=north east] at ([xshift=-0.1cm,yshift=0cm]frame.north east) {\includegraphics[width=1cm]{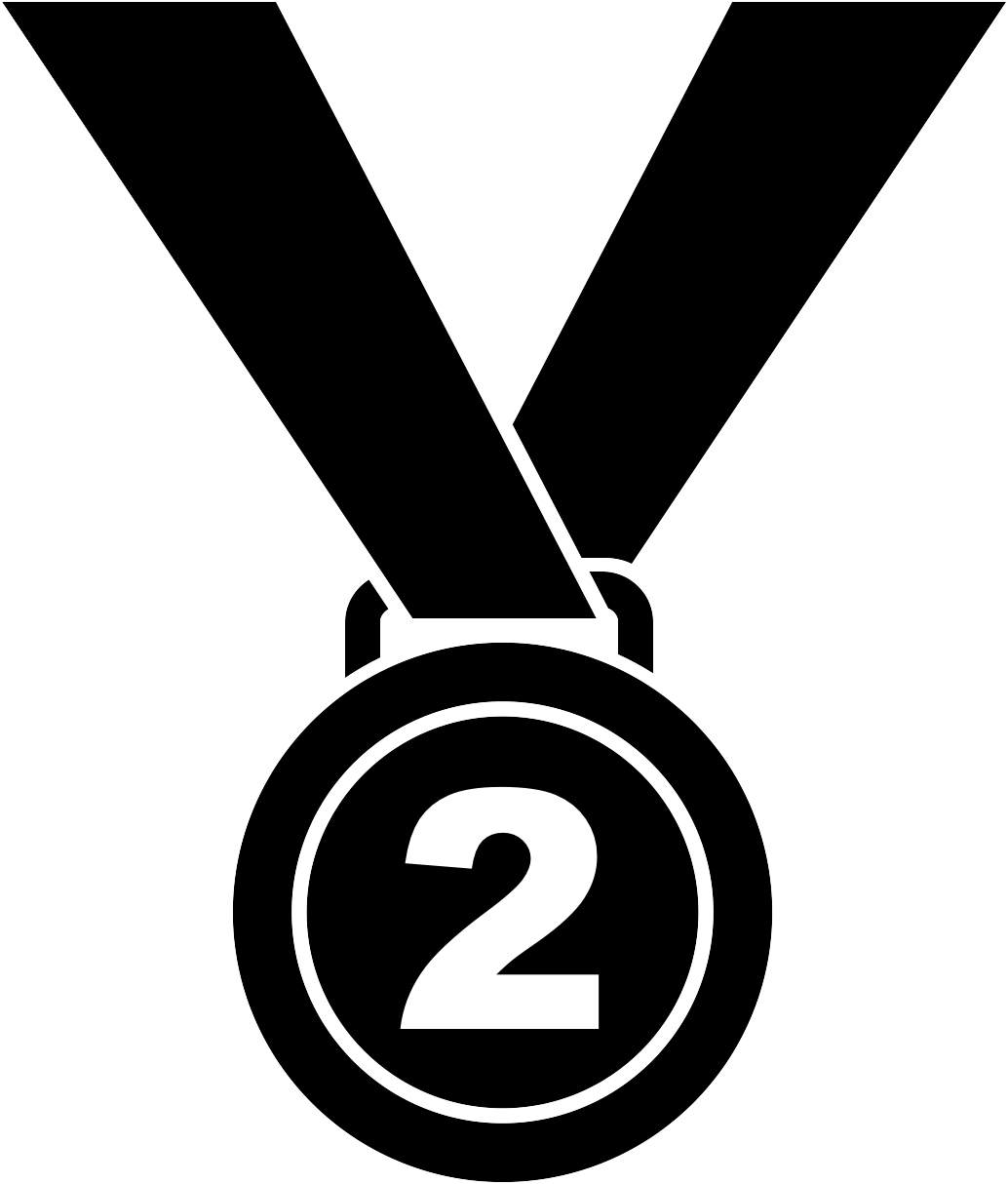}};}]
  \vspace{2mm}
         \includegraphics[width=0.49\textwidth,height=\heightforbox, keepaspectratio]{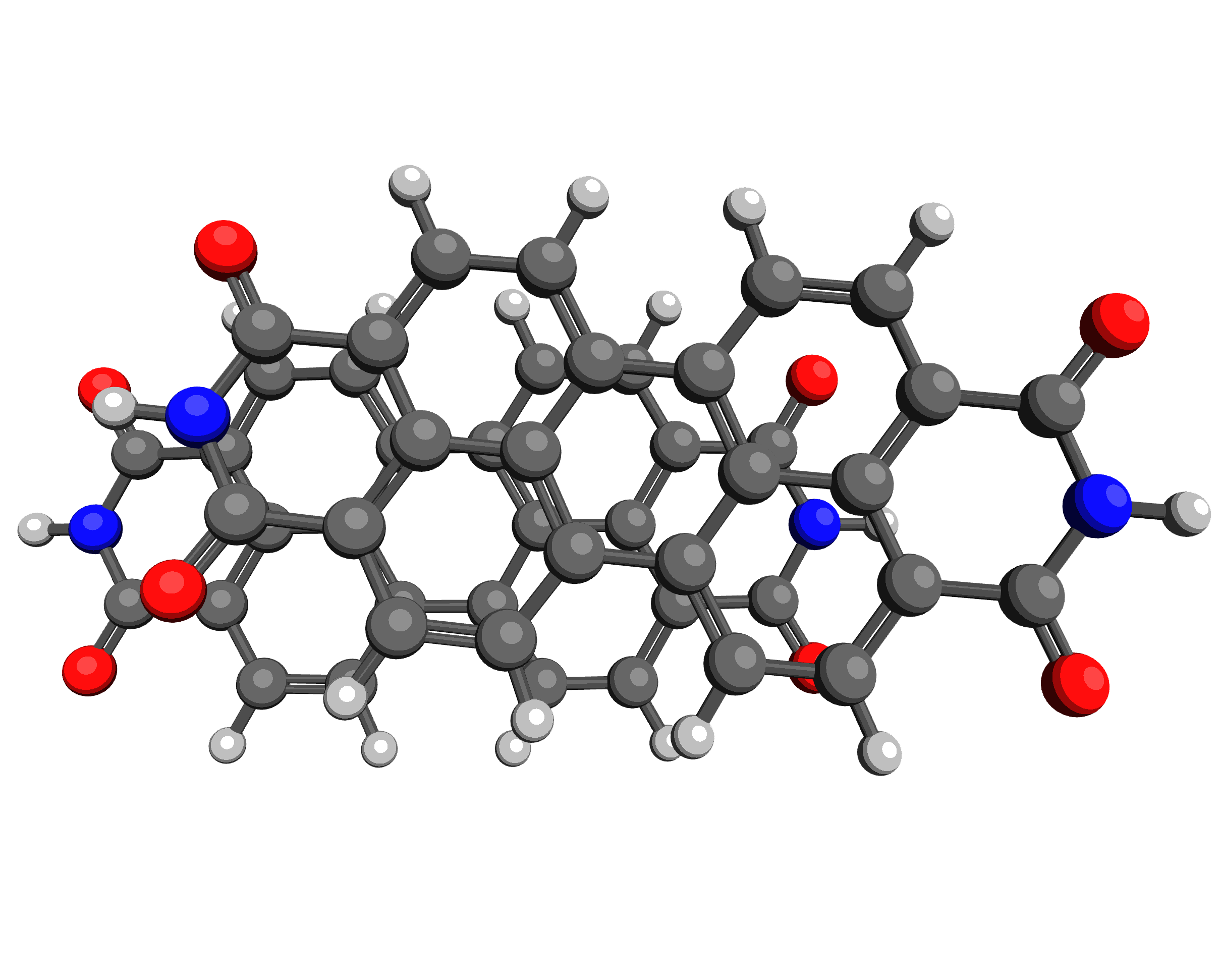}
         \includegraphics[width=0.49\textwidth,height=\heightforbox, keepaspectratio]{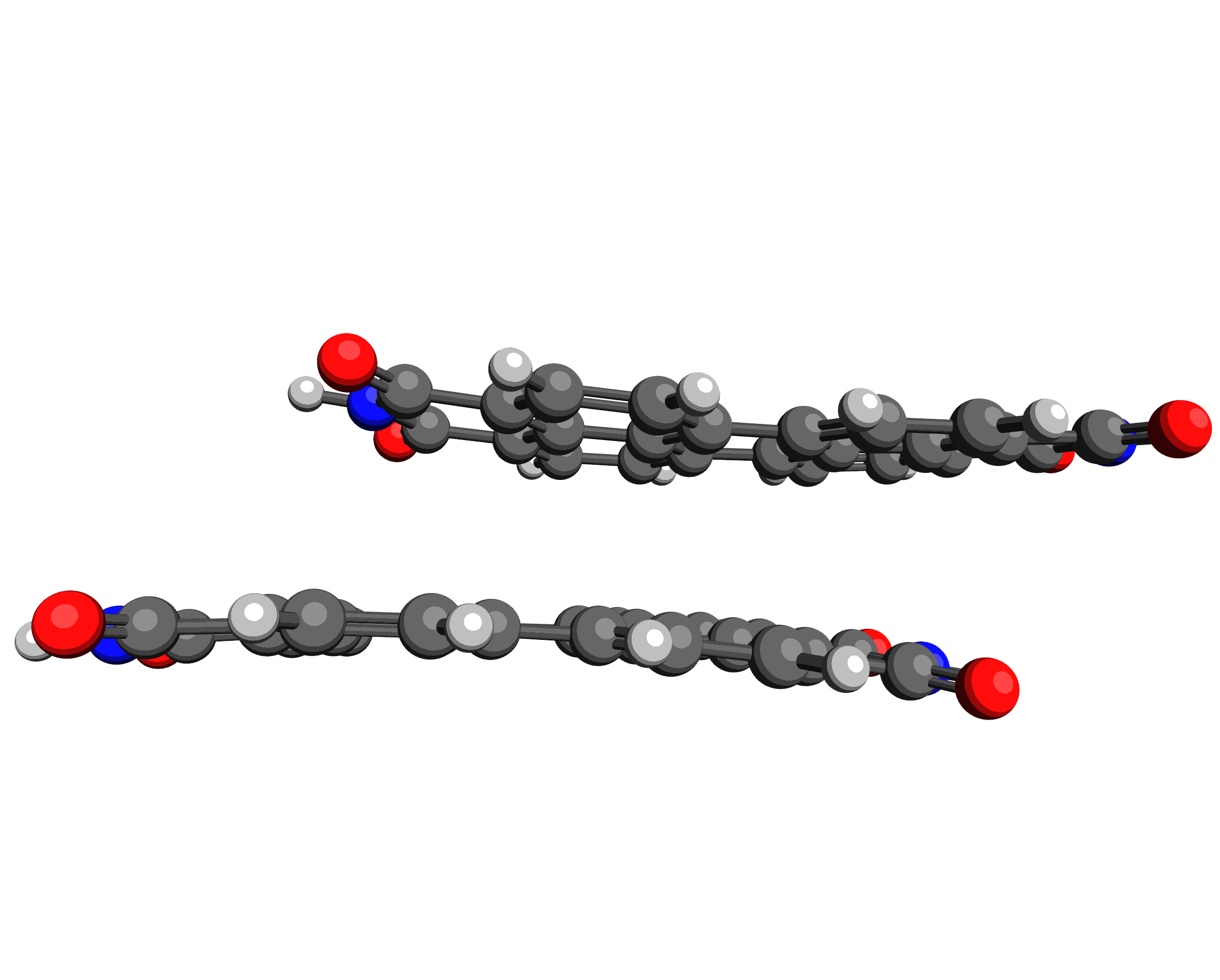}
         \end{tcolorbox}
     \end{subfigure}
     \hspace{0.3cm}
     \begin{subfigure}[b]{0.47\textwidth}
         \centering
         \begin{tcolorbox}[title=(d), enhanced,attach boxed title to top left={yshift=-5.5mm,yshifttext=-1mm},
  colback=white,colframe=plt-cyan,colbacktitle=plt-cyan,
  fonttitle=\bfseries,
  boxed title style={size=small,colframe=plt-cyan}, overlay={
            \node[inner sep=0pt,outer sep=0pt,anchor=north east] at ([xshift=-0.1cm,yshift=0cm]frame.north east) {\includegraphics[width=1cm]{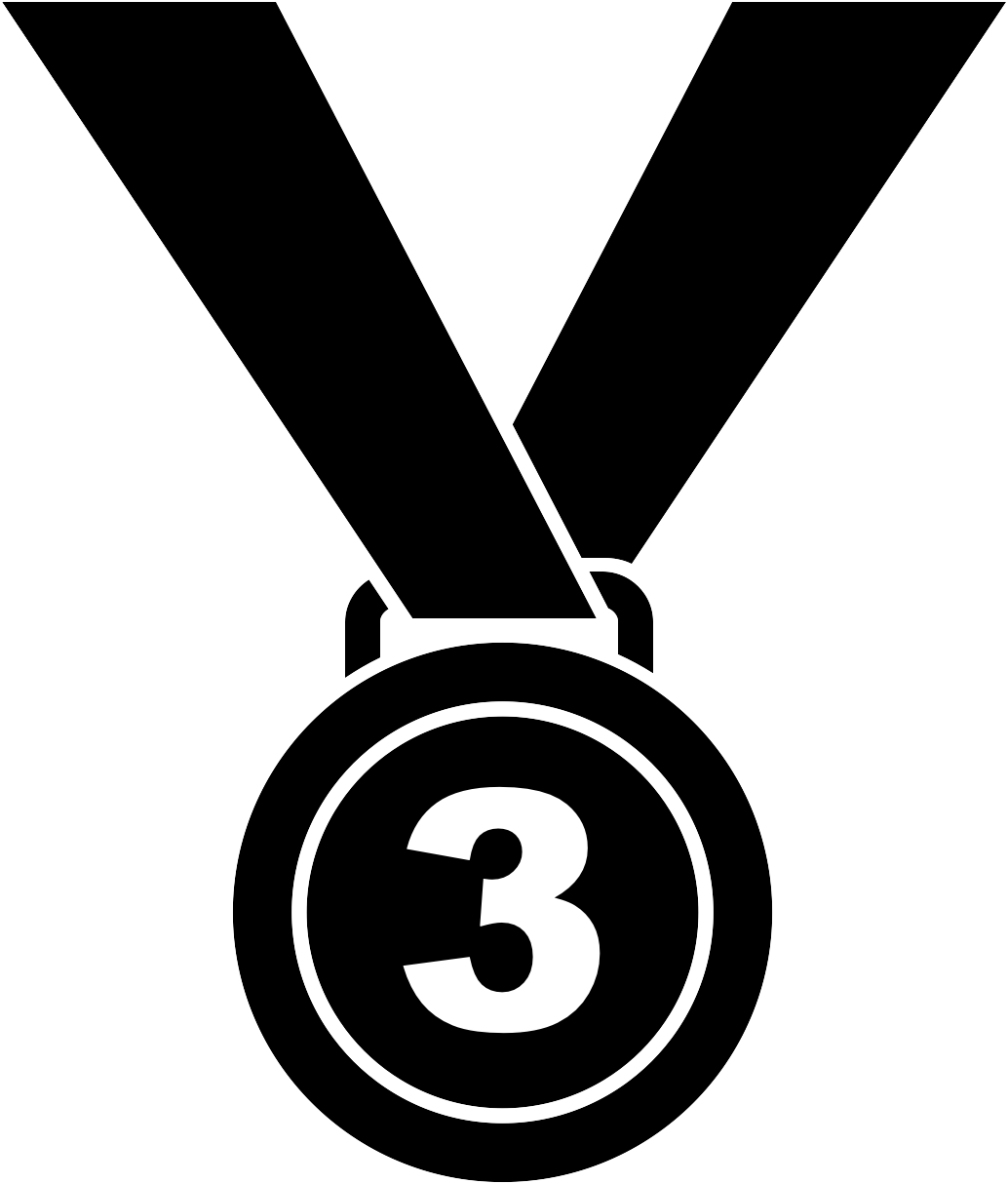}};}]
  \vspace{2mm}
         \includegraphics[width=0.49\textwidth,height=\heightforbox, keepaspectratio]{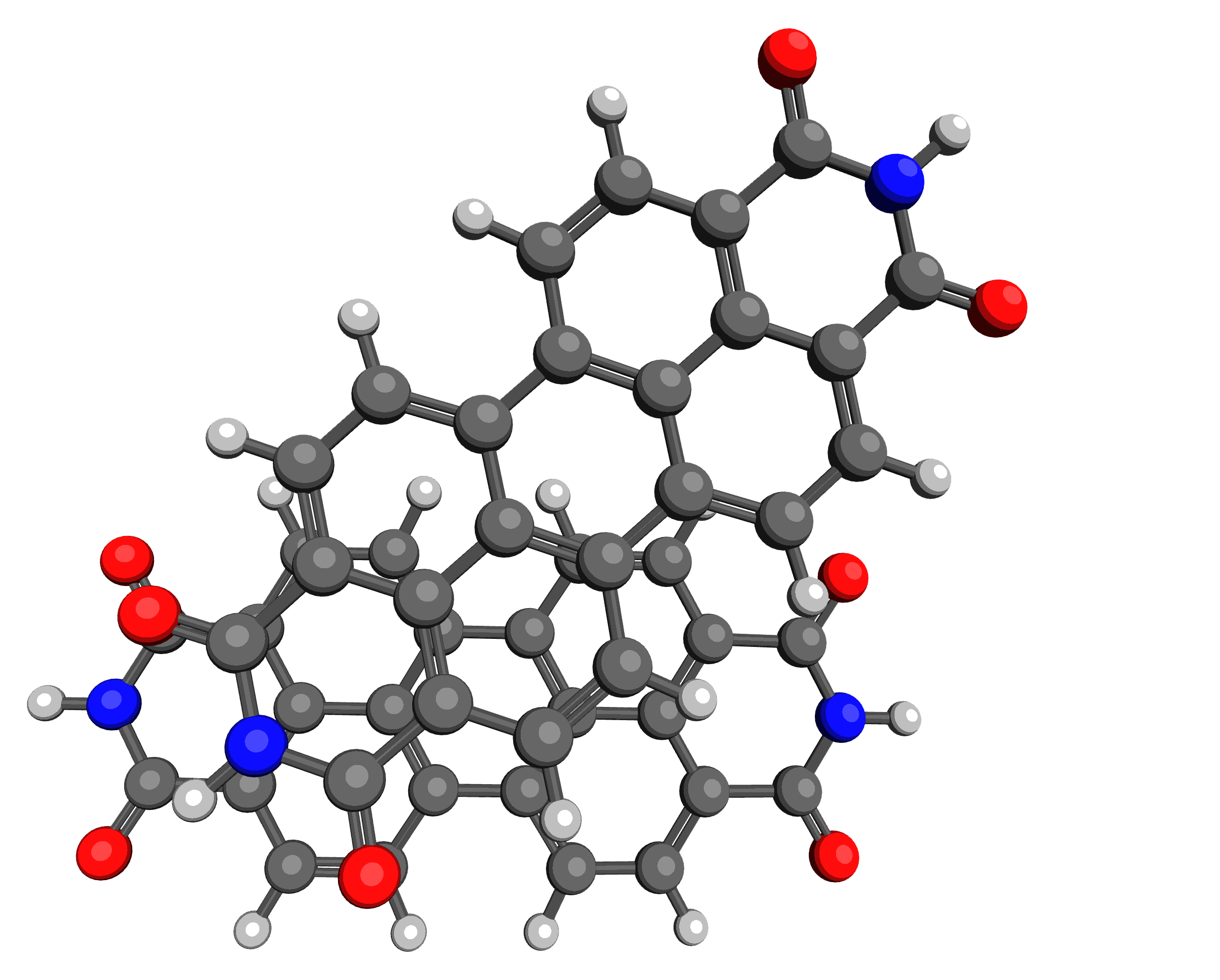}
         \includegraphics[width=0.49\textwidth,height=\heightforbox, keepaspectratio]{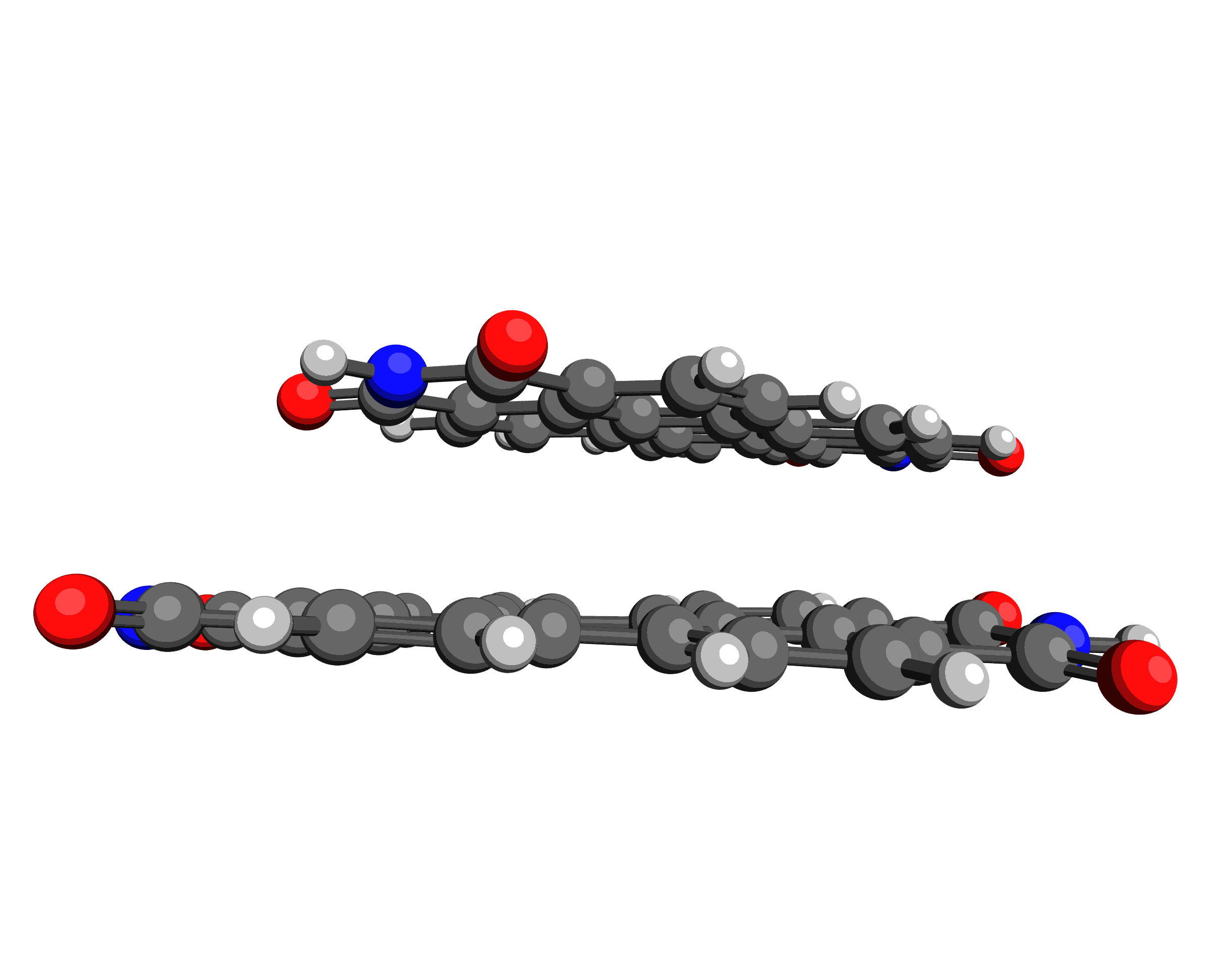}
        \end{tcolorbox}
     \end{subfigure}
     \hfill
    \caption{Generated structures as mean of each cluster. (a) Cluster 1 $\mathbf{PBI-C1}$ with color green: The dimer features a translation in x of \SI{1.46}{\angstrom} and in y of \SI{3.30}{\angstrom}, a z-rotation of \SI{-49.3}{\degree}, curvatures of \SI{3.8e-3}{\per\angstrom\squared} and \SI{1.6e-3}{\per\angstrom\squared} and twisting angles of \SI{10.0}{\degree} and \SI{5.4}{\degree} on monomer $A$ and $B$, respectively. (b) Cluster 2 $\mathbf{PBI-C2}$ with color blue: The dimer features a translation in x of \SI{1.46}{\angstrom} and in y of \SI{1.37}{\angstrom}, a z-rotation of \SI{-24.1}{\degree}, curvatures of \SI{3.6e-3}{\per\angstrom\squared} and \SI{4.7e-3}{\per\angstrom\squared} and twisting angles of \SI{21.6}{\degree} and \SI{14.5}{\degree} on monomer $A$ and $B$, respectively. (c) Cluster 3 $\mathbf{PBI-C3}$ with color chartreuse: The dimer features a translation in x of \SI{3.06}{\angstrom} and in y of \SI{0.72}{\angstrom}, a z-rotation of \SI{-6.2}{\degree}, curvatures of \SI{-18.8e-3}{\per\angstrom\squared} and \SI{19.3e-3}{\per\angstrom\squared} and twisting angles of \SI{7.2}{\degree} and \SI{8.0}{\degree} on monomer $A$ and $B$, respectively. (d) Cluster 4 $\mathbf{PBI-C4}$ with color cyan: The dimer features a translation in x of \SI{1.44}{\angstrom} and in y of \SI{3.79}{\angstrom}, a z-rotation of \SI{43.0}{\degree}, curvatures of \SI{-1.1e-3}{\per\angstrom\squared} and \SI{5.5e-3}{\per\angstrom\squared} and twisting angles of \SI{8.6}{\degree} and \SI{7.2}{\degree} on monomer $A$ and $B$, respectively.}
    \label{fig:clusters}
    \end{center}
\end{figure*}
The individual dimer and monomer structures are in the following named PBI-C$n_{A/B}$, where $n$ denotes the index of the cluster, while subscript $A$ represents the structure of the first monomer and subscript $B$ represents the structure of the second monomer. 
The first cluster ("Cluster 1"), represented in green, exhibits the lowest mean SF rate of \num{1.47e-7}. Its significant attributes include a x-translation of \SI{1.46 \pm 0.54}{\angstrom}, a pronounced y-translation of \SI{3.30 \pm 0.10}{\angstrom} and a strong z-rotation of \SI{-49.3 \pm 9.7}{\degree}.  Curvature (\SI{3.8e-3}{\per\angstrom\squared} and \SI{1.6e-3}{\per\angstrom\squared} as well as twisting ( \SI{10.0}{\degree} and \SI{5.4}{\degree} for monomer $A$ and $B$, respectively) are rather subtle.  

The second cluster ("Cluster 2"), depicted in blue, displays the highest mean SF rate of \num{2.98e-07}, distinguished by notable twist angles in both monomers (\SI{21.6 \pm 5.3}{\degree} for monomer $A$ and  \SI{14.5 \pm 4.1}{\degree} for monomer $B$). This cluster also features significant x and y-translations ( \SI{1.46 \pm 0.39}{\angstrom} and  \SI{1.37 \pm 0.05}{\angstrom}, respectively). The significance of z-rotation \SI{-24.1}{\degree} is reduced due to the large standard deviation of \SI{36.6}{\degree}. Curvatures of both monomers \SI{3.6e-3}{\per\angstrom\squared} and \SI{4.7e-3}{\per\angstrom\squared}) are again subtle.  

The third cluster ("Cluster 3"), given in chartreuse, possesses the second-highest  mean SF rate of \num{2.15e-07} and is primarily characterized by pronounced curvatures in both monomers ( \SI{-18.8 \pm 4.5e-3}{\per\angstrom\squared} for monomer $A$ and  \SI{19.3 \pm 2.0e-3}{\per\angstrom\squared} for monomer $B$). The fact that the signs of the curvatures are opposite, leads to an arrangement in which the molecules are bend in that way that the vertices of the parabola are closest  (see \figurename~\ref{fig:clusters}~(c)). The cluster features a distinct x-translation of \SI{3.06 \pm 0.43}{\angstrom} and slight y-translation of \SI{0.72 \pm 0.16}{\angstrom}, while again, relevance of the  z-rotation is reduced by a large standard deviation of \SI{-6.2\pm 27.9}{\degree}. The twisting angles are subtile ( \SI{7.2}{\degree} and \SI{8.0}{\degree}). 

The fourth cluster (\textbf{PBI-C4}), illustrated in cyan, has a lower mean SF rate of \num{1.67e-07}.The defining characteristics for this cluster include a noticeable x-translation with \SI{1.44 \pm 0.44}{\angstrom} and a strong y-translation with \SI{3.79 \pm 0.16}{\angstrom} as well as a pronounced z-rotation of \SI{43.0 \pm 10.8}{\degree}. Monomer $A$ exhibits virtually no curvature (\SI{-1.1e-3}{\per\angstrom\squared}), while Monomer $B$ displays a slight bend along its length axis (\SI{5.5e-3}{\per\angstrom\squared}). Twisting is subtle (\SI{8.6}{\degree} and \SI{7.2}{\degree} on monomer $A$ and $B$, respectively).

 Notice, that the motifs found for the green cluster and the cyan cluster appear very similar. This is reflected in the  two-dimensional PCA plot in \figurename~\ref{fig:2D_cluster}, in which the  two clusters are nearer to each other than to the others. However, a significant difference consists in the overlapping parts of the individual monomers. In the mean structure of Cluster 1, the outer sections of the aromatic system of both monomers are stacked, while for Cluster 4 we find, that the middle part of one molecule overlaps with the outer aromatic part of the other.

We further calculated adiabatic energies for the $S_{1}$ and $T_{1}$ state of the individual monomers in the discussed "mean" structures in order to evaluate, whether they match the  energy condition for SF, that goes along with the requirement of pronounced electronic couplings. As an ideal case, the singlet state should be slightly higher in energy than twice the energy of the triplet ground state ($2 \Delta E_{T_{1}} - \Delta E_{S_{1}}$), so that the process $S_{0}$ + $S_{1}$ $\rightarrow$ $T_{1}$ + $T_{1}$ is (slightly) exoergic. As the applied optimization routine allows for the individual structural rearrangement of monomers within the dimer system, energies of the molecules within one stack can differ. The obtained results are presented in table \ref{tab:S_1_T_1_CASPT2}. For Cluster 1 and 2, we find exceptionally good energies, in the sense that they clearly fulfill the required energy condition. The monomers of cluster 3, that both show significant contortion, exhibit almost the same energy values as the planar PBI. The energy values of the two monomers of the mean structure of Cluster 4 differ significantly. Deeper inspection of these findings indicate, that already slight twisting leads to drastically improved energy values, while the presence of curvature leads to comparable values as for the planar PBI. As Liu et al.\cite{liu_stringing_2020} noticed, both types of contortions impact the LUMO transition energy, which can be attributed to the overlap of the $\pi$-bond within the PBI's (LUMO), which is amplified by bending, while twisting reduces it.

The significance of our obtained results can be evaluated in the context of recent synthetic work on contorted PBIs. Numerous studies report on the synthesis of twisted PBIs as we have found in this work, that can be achieved through the introduction of ligands and heteroatoms at the bay positions.\cite{queste_synthesis_2010,qiu_suzuki_2006,nowak-krol_progress_2019,nowakkrol_crystalline_2017} Based on the ligand size, the dihedral angle describing the twist can be modulated. Osswald et al.\cite{osswald_effects_2007} reported on the synthesis of a series of twisted PBIs with a variety of halogen atoms attached to the bay positions, exhibiting dihedral angles ranging from 0° to 35°. This spectrum clearly captures the dihedral angle values found in our function-optimized structures.  Also bending of the PBI molecules along the long axis has been reported in Ref. \cite{liu_stringing_2020}, with the degree of curvature fine-tuned by introduction of aromatic ligands of different sizes to the ortho sides of the molecule, acting as "strings" to stabilize the bend. However, a review of the therein presented crystal structures shows significantly more pronounced bending than what we observed for the discussed mean structures, with values between \SI{109e-3}{\per\angstrom\squared} and \SI{163e-3}{\per\angstrom\squared}. Interestingly, Conrad-Burton et al. have experimentally shown that the marked contortion leads to an 2 orders of magnitude increase in the SF rate. They attributed this enhancement to the alignment of excited singlet and triplet state energies in the PBI, determined trough TDDFT calculations.\cite{ConradBurton2019}
In contrast, the maximum curvature observed in our optimized structures  is \SI{-23.4e-3}{\per\angstrom\squared}. To further explore this variance, we calculated the SF rate for the corresponding crystal structure, finding a relatively low  SF rate of \num{2.95e-11}. This suggests, that pronounced bending might be disadvantageous with respect to the electronic coupling strength which we evaluate as a proxy for the  SF rate calculation. 
Finally it should be also mentioned, that the stabilization of two monomer structures bearing different contortion within one aggregate arrangement is synthetically hardly realizable. Nonetheless, we believe that our findings provide a solid foundation for the development of novel PBI-based SF materials. Additionally, the computational methodology outlined herein offers ample opportunities for further refinement and customization of the functionality optimization procedure.

\section{Conclusions}

Within this work, we have introduced a computational strategy to identify structural motifs that exhibit desired properties through a "functionality optimization" procedure. Utilizing a penalty function approach, we formulated an objective function that integrates the ground state energy as well as the inverse of the expression for the chosen "functionality" descriptor. This ensures that the molecular conformation remains physically plausible while enhancing the functionality.  The gradients for both the energy and the descriptor are automatically delivered by the algorithmic differentiation framework. We applied the outlined strategy to discover PBI dimer structures that increase the SF rate, determined by assessing their diabatic couplings.
Through 500 optimizations and subsequent PCA, we identified 4 clusters, each displaying unique structural characteristics. The highest mean SF rate, has been found for Cluster 2, that features a pronounced monomer twist as well as a stacking motif that shows translation both in x- and y- direction as well as rotation around the z-axis. An almost equally high mean SF rate has been found for Cluster 3, that is characterized by pronounced monomer curvature and strong translation along the x axis. The structures of Cluster 1 and Cluster 4  show almost planar structures and share a significant rotation around the z-axis as well as very similar relative translations of the monomers in x- and y-direction. In conclusion, our research lays the groundwork for new PBI-based SF materials as well as establishes  a versatile computational framework for the detection of structural motifs with tailored properties.

\section*{Author Contributions}
J. G.: Writing – Original Draft, Writing  – Review \& Editing, Conceptualization, Investigation, Methodology, Software, Validation, Visualization
A. S.: Investigation, Methodology, Software, Validation
M. I. S. R.: Writing – Original Draft, Writing  – Review \& Editing, Conceptualization, Project Administration, Supervision, Funding Acquisition
\section*{Conflicts of interest}
There are no conflicts to declare'.

\section*{Acknowledgements}
A. S. and M. I. S. R. are thankful for funding by the Bavarian State Initiative "Solar Technologies Go Hybrid". J. G. acknowledges support from Max-Weber-Program scholarship

\bibliographystyle{apsrev4-2}
\bibliography{rsc}

\end{document}


\clearpage
\noindent\begin{center}
    \Large{Functionality Optimization for Singlet Fission Rate Screening in the Full-Dimensional Molecular and Intermolecular  Coordinate Space  - Supporting Information}
    
    \vspace{0.5cm}
    
\large{Johannes Greiner\textit{$^{a,b}$}, Anurag Singh\textit{$^{a,b}$} and Merle I. S. Röhr\textit{$^{\ast a,b}$}}
\end{center}

\small{
\noindent\textit{$^{a}$~Julius-Maximilians-Universität Würzburg, Center for Nanosystems Chemistry, Theodor-Boveri Weg, 97074 Würzburg, Germany. E-mail: merle.roehr@uni-wuerzburg.de}

\noindent\textit{$^{b}$~Julius-Maximilians-Universität Würzburg, Institute of Physical and Theoretical Chemistry, Am Hubland, 97074 Würzburg, Germany. }}

\thispagestyle{empty}
\tableofcontents
\newpage

\setcounter{page}{1}
\begin{landscape}
\section{Analysis of all 500 optimized PBI dimers}
\begin{longtable}{|r|r|r|r|r|r|r|r|r|r|r|r|r|r|r|}
    \caption{500 optimized PBI dimer configurations, ordered from highest to lowest SF Rate. \textbf{id:} index of the original trajectory. \textbf{translation (x,y,z):} Relative intermolecular translation in \si{\angstrom}. \textbf{rotation (x,y,z):} Relative intermolecular rotation in degree (due to symmetry the angle is always positive and lower than \SI{180}{\degree}). \textbf{curvature (A,B):} mean curvature measured along two parallel lines along the long axis of PBI in \SI{1e-3}{\per\angstrom\squared}. \textbf{twist (A,B):} dihedral angle between outer carbons along the long axis of PBI in degrees. \textbf{rate:} SF rate calculated with the overlap approximation. \textbf{$\Delta E$:} AM1 stabilization energy of the dimer with respect to two non-interacting monomers. \textbf{steps:} Optimisation steps until convergence. Maximal step number has been set to 300.}\\
    \hline
    \multirow{2}{*}{rank} & \multirow{2}{*}{id} & \multicolumn{3}{c|}{translation} & \multicolumn{3}{c|}{rotation} & \multicolumn{2}{c|}{curvature} & \multicolumn{2}{c|}{twist} & \multirow{2}{*}{rate} &\multirow{2}{*}{$\Delta E$} &\multirow{2}{*}{steps}\\
    \cline{3-12}
    & & x & y & z & x & y & z & $A$ & $B$ & $A$ & $B$ & & &\\
    \hline
    \endfirsthead
    
    \multicolumn{15}{c}%
    {\tablename\ \thetable{} -- Continued from previous page} \\
    \hline
    \multirow{2}{*}{rank} & \multirow{2}{*}{id} & \multicolumn{3}{c|}{translation} & \multicolumn{3}{c|}{rotation} & \multicolumn{2}{c|}{curvature} & \multicolumn{2}{c|}{twist} & \multirow{2}{*}{rate} &\multirow{2}{*}{$\Delta E$} &\multirow{2}{*}{steps}\\
    \cline{3-12}
    & & x & y & z & x & y & z & A & B & A & B & & &\\
    \hline
    \endhead

    \hline \multicolumn{15}{|c|}{{Continued on next page}} \\ \hline
    \endfoot
    
    \hline
    \endlastfoot
    \num{1} & \num{114} & \num{-0.40} & \num{-0.67} & \num{3.07} & \num{6.1} & \num{6.9} & \num{53.8} & \num{10.6} & \num{-0.7} & \num{24.3} & \num{15.7} & \num{4.06e-07} & \num{0.19} & \num{51}\\
\hline
\num{2} & \num{222} & \num{-0.40} & \num{-0.67} & \num{3.07} & \num{7.7} & \num{7.9} & \num{53.5} & \num{12.5} & \num{-0.8} & \num{26.0} & \num{14.9} & \num{3.99e-07} & \num{0.19} & \num{53}\\
\hline
\num{3} & \num{252} & \num{-0.48} & \num{0.34} & \num{3.40} & \num{174.9} & \num{7.0} & \num{178.7} & \num{-22.6} & \num{23.5} & \num{9.5} & \num{1.8} & \num{3.95e-07} & \num{0.33} & \num{89}\\
\hline
\num{4} & \num{22} & \num{-0.52} & \num{0.29} & \num{3.38} & \num{174.6} & \num{6.4} & \num{178.5} & \num{-20.3} & \num{23.6} & \num{9.0} & \num{2.8} & \num{3.92e-07} & \num{0.32} & \num{77}\\
\hline
\num{5} & \num{31} & \num{0.47} & \num{-0.70} & \num{3.11} & \num{172.0} & \num{8.0} & \num{125.6} & \num{8.9} & \num{1.0} & \num{27.0} & \num{15.3} & \num{3.90e-07} & \num{0.12} & \num{58}\\
\hline
\num{6} & \num{489} & \num{0.50} & \num{-0.66} & \num{3.13} & \num{172.3} & \num{10.6} & \num{124.8} & \num{10.2} & \num{2.8} & \num{28.1} & \num{14.6} & \num{3.90e-07} & \num{0.10} & \num{113}\\
\hline
\num{7} & \num{498} & \num{-0.43} & \num{-0.74} & \num{3.07} & \num{7.8} & \num{7.6} & \num{53.6} & \num{9.6} & \num{-1.1} & \num{28.6} & \num{13.0} & \num{3.88e-07} & \num{0.18} & \num{92}\\
\hline
\num{8} & \num{395} & \num{-0.14} & \num{0.73} & \num{3.13} & \num{1.3} & \num{0.9} & \num{54.6} & \num{9.8} & \num{1.1} & \num{26.3} & \num{15.3} & \num{3.80e-07} & \num{0.18} & \num{59}\\
\hline
\end{longtable}
\end{landscape}

\newpage

\section{Determining an Optimal Value for $\omega$}

In our optimization process, the objective function $L(\textbf{x}) = E(\textbf{x}) - \omega; \log( T_{RP}(\textbf{x})^2)$, where $E(\textbf{x})$ represents the energy and $\vert T_{RP}(\textbf{x}) \vert^2$ signifies the SF rate for the nuclear configuration $\textbf{x}$, necessitates the careful selection of the free parameter $\omega$ to achieve a balance between optimizing energy and SF rate. To this end, we adjust $\omega$ in increments of 0.1, ranging from 0.1 to 10. This adjustment is performed across the same set of 20 randomly translated and rotated ethene dimers, each optimized to minimize $L(\textbf{x})$.

\begin{figure}[ht]
     \centering
     \begin{subfigure}[b]{0.45\textwidth}
         \centering
         \includegraphics[width=\textwidth]{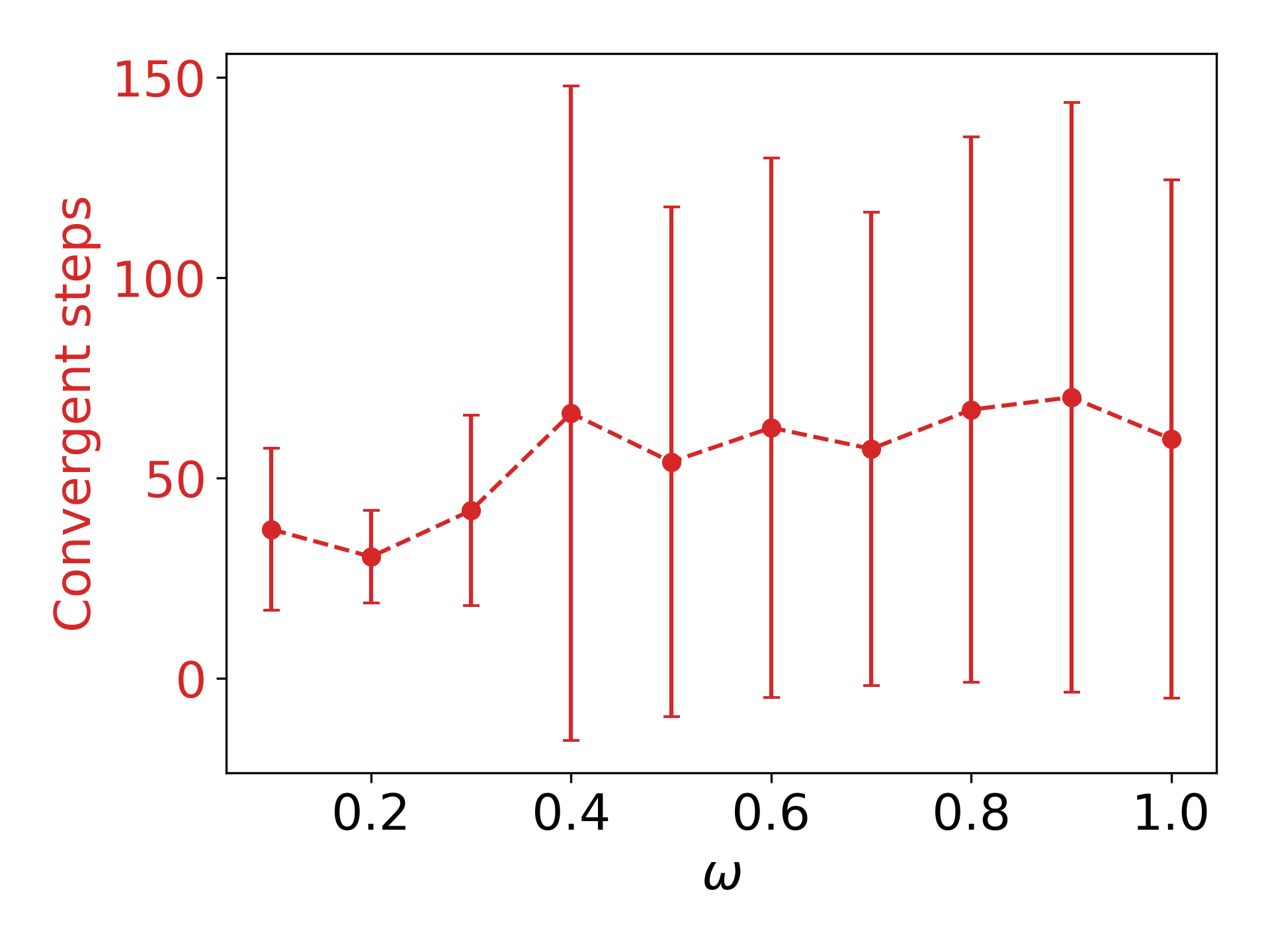}
         \caption{}
         \label{fig:omega_scan_converegence}
     \end{subfigure}
     \hfill
     \begin{subfigure}[b]{0.45\textwidth}
         \centering
         \includegraphics[width=\textwidth]{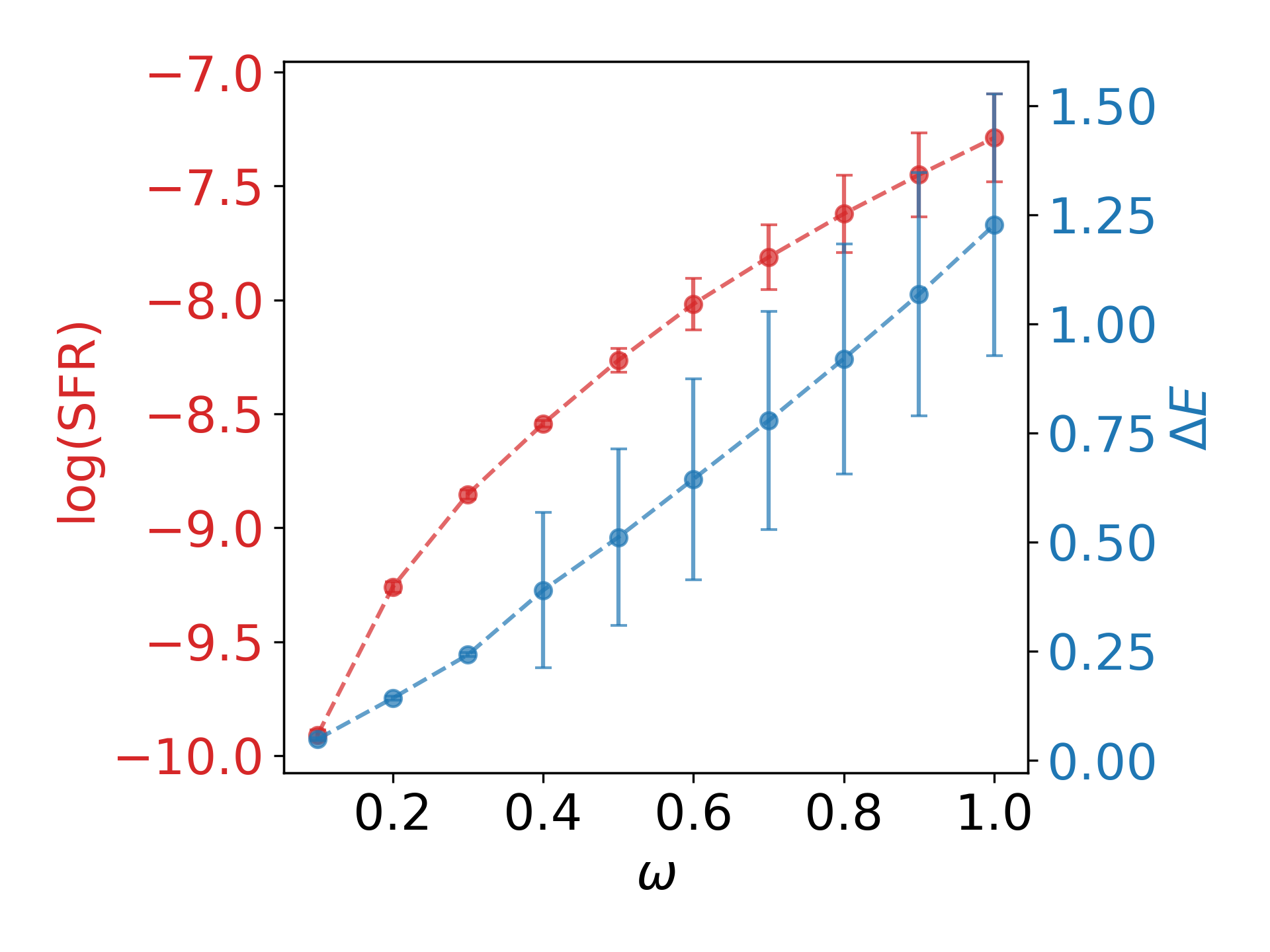}
         \caption{}
         \label{fig:omega_scan_sfr_dE}
     \end{subfigure}
        \caption{Optimization procedure with varying $\omega$ from 0.1 to 1.0 in steps of 0.1 for the same 20 random ethene dimers each.}
        \label{fig:omega_scan}
\end{figure}

The average and standard deviation of the convergent steps, along with the SF rate (SFR) and energy for each $\omega$, are depicted in \figurename~\ref{fig:omega_scan}. The graphical representation in \figurename~\ref{fig:omega_scan_sfr_dE} illustrates a strong positive correlation between SF rate and energy, complicating the choice of an ideal $\omega$. Since our goal is to minimize energy while maximizing SF rate, we select the value of $\omega$ that maximizes the hysteresis between these two parameters as optimal. Consequently, $\omega = 0.3$ has been identified as the preferred choice. Moreover, values exceeding this threshold tend to exhibit greater instability in terms of convergence steps and energy, as indicated by their standard deviations. 

